\documentclass{article}

\usepackage{arxiv}

\usepackage[utf8]{inputenc} 
\usepackage[T1]{fontenc}    
\usepackage{hyperref}       
\usepackage{url}            
\usepackage{booktabs}       
\usepackage{amsfonts}       
\usepackage{nicefrac}       
\usepackage{microtype}      
\usepackage{lipsum}		
\usepackage{graphicx}
\usepackage{natbib}
\usepackage{doi}
\usepackage{float}

\title{Selective pruning and neuronal death generate heavy-tail network connectivity}

\author{ \href{orcid.org/0000-0002-3511-6439}{\includegraphics[scale=0.06]{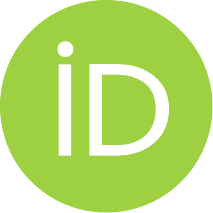}\hspace{1mm}Rodrigo Kazu Siqueira}\thanks{Kazu RS is also affiliated to the INSIGNEO Institute, University of Sheffield; \href{http://neuralpathways.uk}{Neuralpathways website} \& \href{https://metabio.netlify.app/}{metaBIO website} }  \\\ Wellcome Sanger Institute \\
Cambridge, England, UK \\
	\texttt{rs47@sanger.ac.uk} \\\label{key}
	\And
	\href{https://orcid.org/0000-0001-9519-4909}{\includegraphics[scale=0.06]{orcid.pdf}\hspace{1mm}Kleber Neves} \\
	Institute of Medical Biochemistry\\
	Federal University of Rio de Janeiro, Brazil\\
	\texttt{kleber.neves@bioqmed.ufrj.br} \\
 	\And
	\href{https://orcid.org/0000-0000-0000-0000}{\includegraphics[scale=0.06]{orcid.pdf}\hspace{1mm}Bruno Mota} \\
	Physics Institute\\
	Federal University of Rio de Janeiro, Brazil\\
	\texttt{bruno@if.ufrj.br}
}

\renewcommand{\shorttitle}{Kazu \textit{et al}. Selective pruning and neuronal death generate heavy-tail network connectivity}

\hypersetup{
pdftitle={Kazu et al. Selective pruning and neuronal death generate scale-invariant neuronal networks},
pdfsubject={q-bio.NC, q-bio.QM},
pdfauthor={Kazu RS., Neves, K., Mota, B.},
pdfkeywords={SCALE-FREE, SYNAPTIC PRUNING, BRAIN DEVELOPMENT, NEURONAL DEATH, HEAVY-TAIL CONNECTIVITY, FEED-FORWARD NETWORKS, PREFERENTIAL DETACHMENT, POWER LAW},
}

\begin{document}
\maketitle

\begin{abstract}

From the proliferative mechanisms generating neurons from progenitor cells to neuron migration and synaptic connection formation, several vicissitudes culminate in the mature brain. Both component loss and gain remain ubiquitous during brain development. For example, rodent brains lose over half of their initial neurons and synapses during healthy development. The role of deleterious steps in network ontogeny remains unclear, yet it is unlikely these costly processes are random. Like neurogenesis and synaptogenesis, synaptic pruning and neuron death likely evolved to support complex, efficient computations. In order to incorporate both component loss and gain in describing neuronal networks, we propose an algorithm where a directed network evolves through the selective deletion of less-connected nodes (neurons) and edges (synapses). Resulting in networks that display scale-invariant degree distributions, provided the network is predominantly feed-forward. Scale-invariance offers several advantages in biological networks: scalability, resistance to random deletions, and strong connectivity with parsimonious wiring. Whilst our algorithm is not intended to be a realistic model of neuronal network formation, our results suggest selective deletion is an adaptive mechanism contributing to more stable and efficient networks. This process aligns with observed decreasing pruning rates in animal studies, resulting in higher synapse preservation. Our overall findings have broader implications for network science. Scale-invariance in degree distributions was demonstrated in growing preferential attachment networks and observed empirically. Our preferential detachment algorithm offers an alternative mechanism for generating such networks, suggesting that both mechanisms may be part of a broader class of algorithms resulting in scale-free networks.

\end{abstract}


\pagebreak

\section{Introduction}

The normal process of ontogeny of the human brain is not entirely constructive. To a large extent, brain development is destructive, removing parts - neurons and synapses - generated in excess beforehand \citep{abitz2007excess, petanjek2011extraordinary}. It has been known for a long time that there is widespread neuronal death and synaptic pruning in the development of the nervous system \citep{hamburger1990naturally,changeux1976selectivePruning}.
For instance, studies in rodents have shown that the number of neurons at birth is almost double that of the adult, with half of the initial number of neurons being eliminated by programmed cell death in the first weeks of life \citep{bandeira2009changing}. While this might seem like an incredibly wasteful way of building brain circuitry, this process is thought to be a natural part of brain development and not a pathology \citep{clarke1985neuronalDeath}. Inhibition of this regulated cell death causes differences in the number of neurons, size and shape of brains in mammals \citep{haydar1999apoptosisBrainSize}. 

A similar phenomenon happens with synapses, albeit in a different time scale. The pruning of the initial exuberance of synapses starts early and goes on until adulthood, being necessary for normal development \citep{huttenlocher1979synapticDensity, stiles2010basics}. The timing and intensity of pruning, however, vary across different brain regions - e.g., sensory systems prune early as opposed to the frontal cortex that is late maturing, in this sense \citep{elston2009spinogenesisPruning}. Interestingly, this is the period thought to be when the brain is at peak plasticity \citep{craik2006lifespan}.
	
Which neurons and synapses survive and which ones are eliminated?

It is well-known that the elimination of neurons and synapses is activity-dependent \citep{clarke1985neuronalDeath, snider1992axotomy} and that synaptic pruning depends on microglial activity \citep{paolicelli2011synaptic}. These resident immune cells are thought to make direct and ephemeral connections with synapses in a way that is quantitatively proportional to basal neuronal activity \citep{wake2009resting}. Furthermore, Schafer \textit{et al.} showed in 2012 that microglia engulfs considerably more synapses of the less active cells (Tetrodotoxin-treated) as compared to untreated ones and that this tends to happen at the presynaptic terminals and relies on the microglia-specific phagocytic signalling pathway, complement receptor 3(CR3)/C3 \citep{schafer2012microglia}. 

Alterations in number/density of synapses have been shown to be associated with conditions such as autism \citep{thomas2016over, teter2024crispri}, glaucoma \citep{rosen2010role, howell2011molecular}, Huntington’s disease \citep{wilton2023microglia}, schizophrenia \citep{yilmaz2021overexpression} and are distinguishing features of Alzheimer's disease (AD) \citep{hong2016complement, lui2016progranulin}, amongst others \citep{hong2016new, scott2023convergent}. Notably, it is thought that the same mechanisms that prune excess synapses during development are inappropriately activated in AD \citep{hong2016complement}.

The strengthening or weakening of synapses is thought to be regulated through Hebbian mechanisms, as in "neurons that fire together, wire together" \citep{hebb1949organization}. These are the basis, for example, for the plasticity in neuronal circuits of the hippocampus \citep{bastrikova2008pruningHp} and for the experience-dependent refinement of visual circuits in the retina \citep{zhou2004pruningLTD, butts2007burstHebb, zhang2012hebbVisual}. Similarly, in the initial stages of development, neurons that are being regularly activated by peripheral innervation are the ones likely to survive. This mechanism for survival is thought to involve retrograde transport of growth factors, such as NGF and BDNF and other factors released by glial cells  \citep{ghosh1994requirement}. Since the axonal path of every neuron cannot be specified in advance due to the sheer number of neuronal cells in the human brain \citep{herculano2009human}, this is thought to be a solution to the problem of exactly finding an innervation target: project neurons in excess and remove the ones that fail to reach a target. It seems plausible that this "wasteful exuberance" is a property that was selected for, a consequence of a biophysical principle to minimize the total amount of axonal wiring \citep{raj2011wiring}.
	
Even though a lot is known about the function and even possible molecular and cellular mechanisms for neuronal death and synaptic pruning, the effect of these events, their timing and their order on the structure and functioning of the resulting neuronal network has not been explored.
 
\subsection{Scale-invariance in neuronal networks} 

Our current understanding of anatomical connectivity can be traced to the late 19th century, with the postulation of the neuron doctrine. It has been since assumed that the brain is composed primarily of individual neuronal cells that are interconnected via synapses. Under this premise, the brain is a system consisting of clearly defined interconnected elements, and its functioning is determined by its observable connectivity. Mathematically, there is natural representation of such structure as a directed network of interacting nodes, to which one can apply quantitative tools and the theoretical framework borrowed from network science. This has been a fruitful approach to investigate brain activity and, far less often, its ontology. This relatively new field is called network neuroscience \citep{bassett2017network}.

Functional brain connectivity studies have revealed nonrandom topological features. For example, numerous studies have identified a highly clustered cortical network characterized by short average path lengths between nodes and a small-world topology \citep{bullmore2009complex, bassett2017small} (for a comprehensive review, see \citep{van2016comparative}). Additionally, significant evidence suggests that the brain network is divided into distinct communities \citep{puxeddu2020modular}, interconnected by metabolically expensive hub nodes \citep{crossley2014hubs}, forming what is known as a 'rich club' \citep{van2011rich}. 

However, when modelling the nervous system as a network, nodes and edges can defined at different levels, from lower (cellular or even molecular) to higher ones (large populations of cells or well-defined anatomical regions). This raises concerns that range from the experimental technique utilised for data collection to the problem of node definition \citep{stanley2013defining}. Modelling the network at the cellular level adds another difficulty since experimentally mapping the details of neuronal connectivity is a significant technical challenge \citep{helmstaedter2013cellular}. Only recently has large-scale neuronal connectivity been investigated at cellular resolution in whole organisms. These findings showed heavy-tail connectivity in the partial connectome of the \textit{Drosophila} central brain and its optic medulla \citep{scheffer2020connectome, takemura2013visual}. 

Specifically, scale-free connectivity is also found in the connectome of the roundworm \textit{Caenorhabditis elegans} \citep{white1986structure, towlson2013rich} and the sensory-motor pathways of the annelid \textit{Platynereis} \citep{randel2014neuronal, lynn2024heavy}. In vertebrates, evidence for heavy-tail connectivity was found in the mouse retina both for number and size distribution of physical contacts in between neurons \citep{stringer2019high, lynn2024heavy}. These studies collectively highlight the prevalence of heavy-tail connectivity in neuronal networks across different species.

The presence of scale-free connectivity suggests that a small fraction of neurons are significantly more connected than the majority, indicating a non-random organization that likely plays a crucial role in network dynamics and information processing within these neural circuits \citep{barabasi2009scale}. We can infer that neuronal networks will be robust to random loss of neurons, since neuronal loss with ageing occurs, at a slow but steady rate, throughout all adulthood \citep{mortera2012age}. This suggests that at a low level, it is plausible that neuronal networks are scale-invariant, with a degree distribution (the distribution of the number of connections of each neuronal cell) following a power law.

Known mechanisms to generate scale-free networks involve preferential attachment \citep{barabasiAlbert1999, bianconi2001competition, bell2017network}. Nodes in the network connect preferentially to the nodes with the highest intrinsic fitness or connectivity. However, this class of algorithms is not plausible biologically: neurons would require global information on the fitness and/or connectivity of other neurons \citep{nicosia2013co}.

To explore the effects that the aforementioned ontogenetic events might have on the low-level organization of neuronal networks, here we take as our object of study an idealized cortical circuit \citep{douglas1989canonical} and perform our investigation through computational modelling. We devised a modular, evolving, and directed network of 50,000 neurons and 5,000,000 synapses. This was thought to roughly double both the number of cells in cortical columns that were reported to have 19.000 cells in mice \citep{meyer2010number} and the 21.739 cells sampled in the partial connectome of the \textit{Drosophila} central brain \citep{scheffer2020connectome}. We found this number reasonable especially due to the nonuniformity found in primates that have no fixed number of neurons under 1 mm$^2$ of the cortex \citep{herculano2008basic} but that would have a higher density than any rodent of the same brain mass \citep{gabi2010cellular}.

We hypothesized that both the selective death of neurons and the pruning of synapses would lead to significant changes in the topology of neuronal networks. Specifically, we expected these changes to enhance their robustness, reduce average path lengths and increase overall clustering. Our model builds upon fundamental assumptions derived from neurobiology: (1) neuronal death and synaptic pruning happen on a large scale during brain development, (2) their occurrence is selective, influenced by connectivity patterns, (3) neurons primarily access local connectivity information, and (4) the network exhibits a feed-forward structure with a preferred direction for most connections.

By combining these elements, we aim to provide insights into the interplay of brain development and network organization.

\section{The model}\label{Methods}

 \subsection*{Definition of the Network}

We model the network of neurons as graph $G = (V, E)$, where $V$ is the set of neurons (vertices); and $E$ is the set of synapses (edges). Connections between neurons are directed but binary - activity is not explicitly modelled. Time is discrete in the model. The evolution of the network over time is defined as follows:

\begin{itemize}
\item \emph{Initial condition:} start with a network $G$ with $N_{n}$ ordered nodes and $N_{e}$ randomly added edges, with the direction of each edge chosen independently according to the percentage of feed-forward edges;

\item \emph{Neuronal death stage (ND):} For each node $n$, with probability $D(n)$, remove the node and all its edges. The probability $D_{n}$ is calculated according to equation (\ref{eq:death}) (see below). After elimination, the same amount of nodes and edges removed is randomly added back to the network, restoring the number of nodes and edges to $N_n$ and $N_e$ at the end of each step.

\item \emph{Synaptic pruning stage (SP):} For each existing edge $E(i,j)$, with probability $P(i,j)$ (equation (\ref{eq:pruning}) below), remove the edge. Contrary to the neuronal death stage, no edges or nodes are added back to the network in this stage. The reasoning here is the longer duration of the synaptic pruning, which extends through adulthood in humans.
\end{itemize}

In the model, the ND stage represents the short post-natal period of intense cell death occurring in parallel with cortical neurogenesis, where the number of neurons goes through large variations but ends about the same after the process \citep{wong2019developmental}. The SP stage represents the pruning of synapses in the course of development (through adulthood in humans, see \citep{nagappan2023molecular}).

Death and pruning probabilities are calculated as follows:
	
\begin{equation} \label{eq:death}
D(n) = A \exp\left(-k \frac{d_{IN}(n)  N}{\sum_{m}d_{IN}(m)}\right)
\end{equation}

\begin{equation} \label{eq:pruning}
P(i,j) = A \exp\left(-k \frac{d_{IN}(j) d_{OUT}(i) N^2}{\sum_{m} d_{IN}(m) \sum_{m} d_{OUT}(m)}\right)
\end{equation}

where $d_{IN}(n)$ and $d_{OUT}(n)$ are, respectively, the incoming and outgoing degrees of node $n$; and $N$ is the number of nodes. The parameters $A$ and $k$ are used to regulate the rate of pruning in the simulations. The formula for neuronal death is thought to reflect activity-dependent survival; activity in a neuron is assumed to be approximately proportional to a number of incoming edges, through which the neuron would receive growth factors through anterograde transport \citep{tonra1998axotomy, von2000expression}. In the case of synaptic pruning, we used the Hebbian learning mechanism of preserving the synapses which fire synchronously – to save computational costs, for an edge going from node $i$ to node $j$, this is approximated by the product of the number of incoming edges of $j$ and the number of outgoing edges of $i$.

Notice that the formulas take the degree of each node into account. This is the degree relative to the previous step, since degrees are altered by the removal of edges themselves, this could lead to a cascading effect if we used the updated values (i.e. a "hot" update). Nodes/edges are removed only after all probabilities are calculated.
	
\begin{figure} [H]
	
	\begin{center}
		\includegraphics[width=1\textwidth]{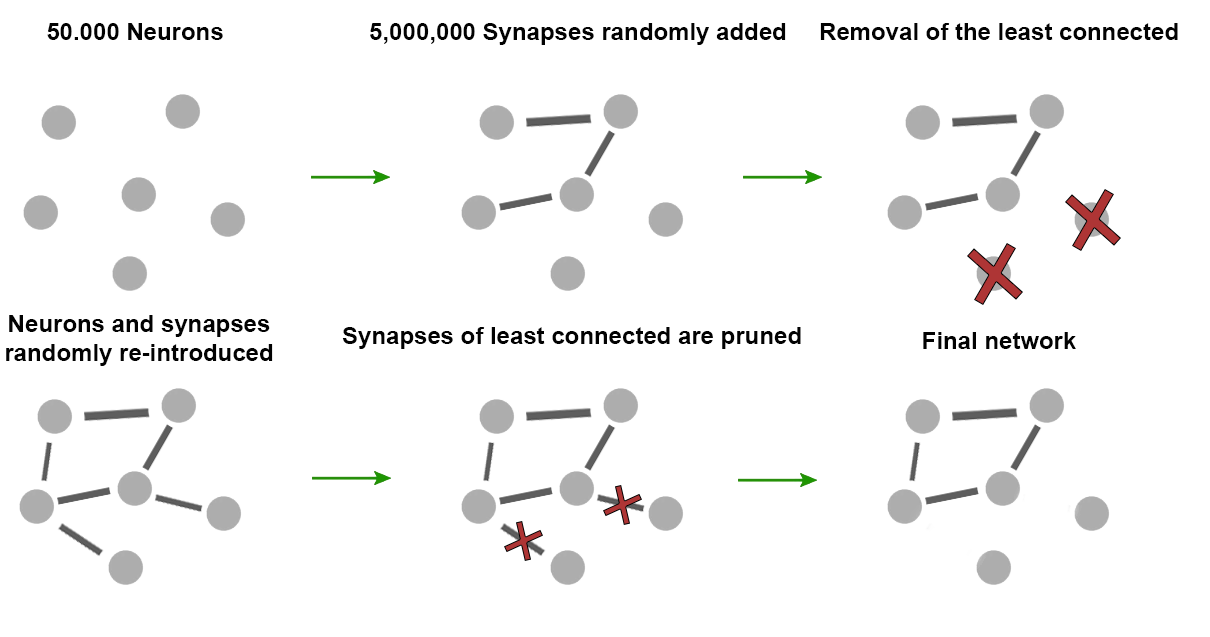}
		\caption{\label{fig:graph_elements} Main steps of our modular, evolving, and directed network graph.}
	\end{center}
	
\end{figure}

Typically, in our simulations, $N_{n}$ is 50,000 and $N_{e}$ is 100,000 (Figure \ref{fig:graph_elements}), and the number of iterations in the ND stage is set at 500. This was determined empirically: we noticed that after 500 iterations, the degree distribution seems stable. The number of iterations in the SP stage is determined naturally since the network is eventually destroyed by the removal of edges.

\subsection*{Implementation \& Data Analysis}

	The model was implemented in Python 3 \citep{van1995python}, using the \textit{igraph} \citep{csardi2006igraph} and \textit{SciPy} \citep{bressert2012scipy} libraries. All posterior analysis and graphing were performed in Python, using \textit{matplotlib} \citep{hunter2007matplotlib} and \textit{Seaborn} \citep{Waskom2017}. The goodness of fit tests for the degree distributions was performed in accordance with the recommendations in \citep{clauset2009powerlaw}.

\section{Results}

 The base model is generated using the criteria for neuronal death and synaptic pruning we chose \textit{a priori} to resemble biological events of brain development (see Figure \ref{fig:graph_elements} in the Methods section for a detailed description).

We then compared the outputs of our model with a number of conditions. Namely, we wanted to measure the direct effects of the selective death and the selective pruning on the degree distributions after the model runs for the iterations needed for the ND and SP stages. First, we tested whether selective death is an essential factor. To test that, we ran the same simulation, the only change being the criteria for neuronal elimination in the death stage: we kept the same death rate as in the base model, but instead of the probability of death being inversely related to the degree, their dying probabilities were random. Furthermore, we performed a similar change for the probability of edges being pruned from the network.

\subsection{Degree distributions}

\begin{figure} [H]
	
	\begin{center}
		\includegraphics[width=1\textwidth]{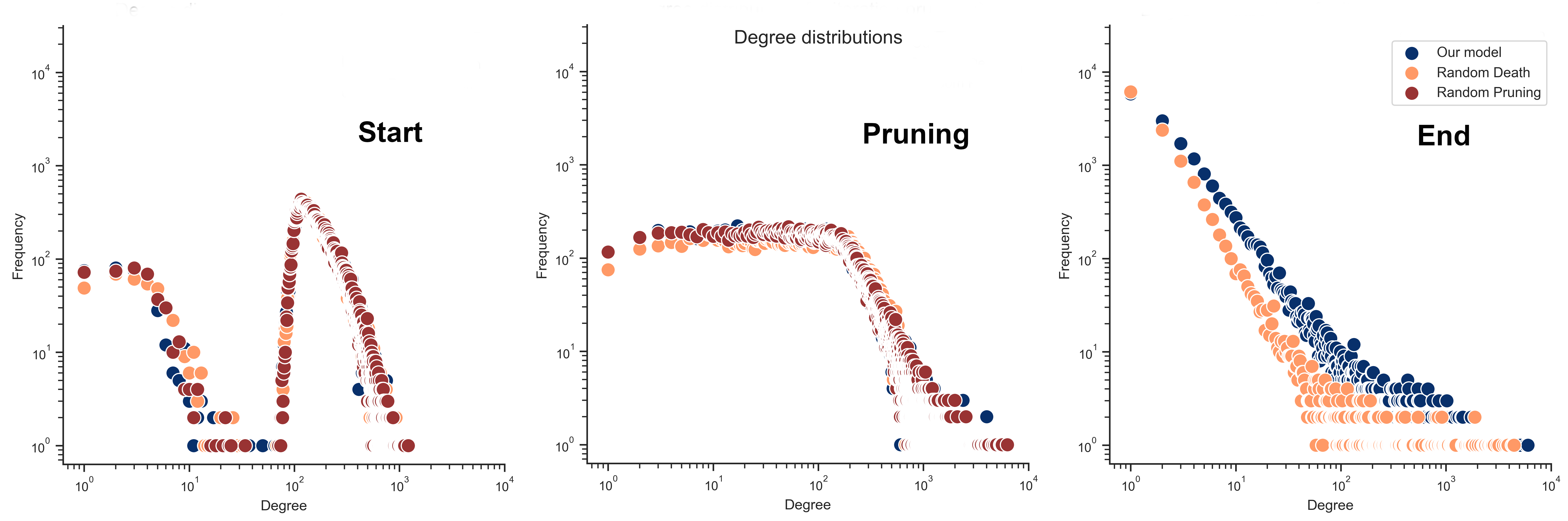}
		\caption{\label{fig:main_result_comparativesyn100} Distribution of the number of synapses (degree) of each neuron (node) in the network. The panels present the progression of the degree distribution in our model and colours stand for the different conditions tested, with blue being our base model with selective death and pruning, orange being when only the death is random and red when only the pruning is random.}
	\end{center}
	
\end{figure}

In figure \ref{fig:main_result_comparativesyn100} we show the degree distributions for our model as it progresses through its successive stages.  During the neuronal death stage, the initial degree distribution converges to a combination of constant degree distribution for low degrees and a fat-tailed distribution for higher degrees

during the death stage starts with neurons of different degrees spread across the distribution and progresses towards a state where the rightmost portion of the distribution seems to have a heavy tail characteristic of power laws (Figure \ref{fig:main_result_comparativesyn100} - "Pruning" panel). Furthermore, our model (in blue) starts to show a more structured decline than the ones with random death (in orange). This distribution is stable (Figure \ref{fig:main_result_comparativesyn100} - "End" panel), up until the point where the number of neurons becomes too low. In Figure \ref{fig:main_result_comparativesyn100} - "End" panel, we see the cumulative degree distribution as the simulation continues to the pruning stage. 

\begin{figure} [H]
	
 	\begin{center}
 		\includegraphics[width=1\textwidth]{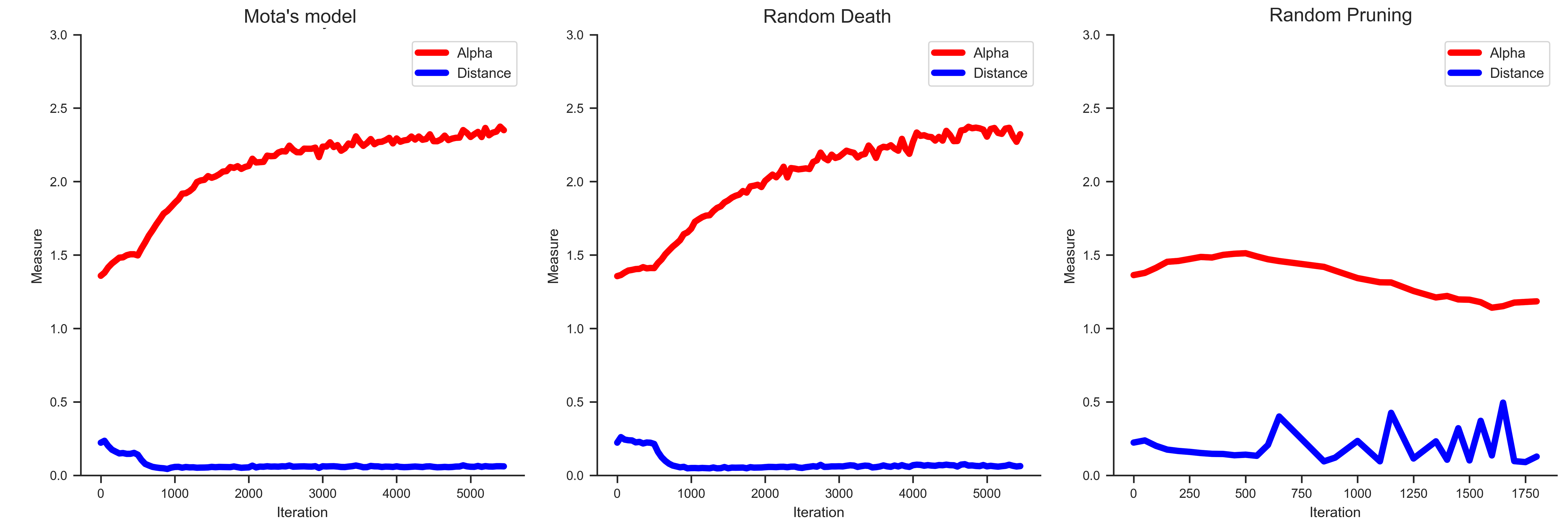}
 		\caption{\label{fig:Alpha_Dsyn100} The evolution of the exponent alpha and the Kolmogorov distance D as the simulation evolves.}
 	\end{center}
	
 \end{figure}

We then evaluated if, as the iterations went on, the distributions progressed to a power-law distribution utilising the Kolmogorov-Smirnovtest (K-S) \citep{cong2020comprehensible}  test. The evolution of the exponent alpha and the Kolmogorov-Smirnov distance D can be seen in figure  \ref{fig:Alpha_Dsyn100}.  At iteration 5400 we have a D of 0.06 between the empirical distribution and a power law distribution with exponent 2.37.  We took snapshots of the network and calculated the statistics for all 50 iterations of all of our generated networks. As the network evolves D gets smaller, pointing towards both our model and the network that was generated with random death being power laws (D = 0.06, alpha = 2.27, at 5400 iterations for the network generated with random death). The same cannot be said of the network generated with selective death and random pruning where the D varies in a less predictable fashion.  Substituting the selective pruning for random pruning in the model makes the network slowly lose its edges randomly, thinning out until it is completely disconnected. It seems that not only a pruning stage, but a particular kind of pruning criteria is essential to generate the scale-free behaviour in the network model. The K-S test results clearly demonstrate that our model outperforms both random death and random pruning in maintaining a power-law distribution in the network. 

\subsection{Complex network statistics}
  
\begin{figure} [H]
	
	\begin{center}
		\includegraphics[width=0.88\textwidth]{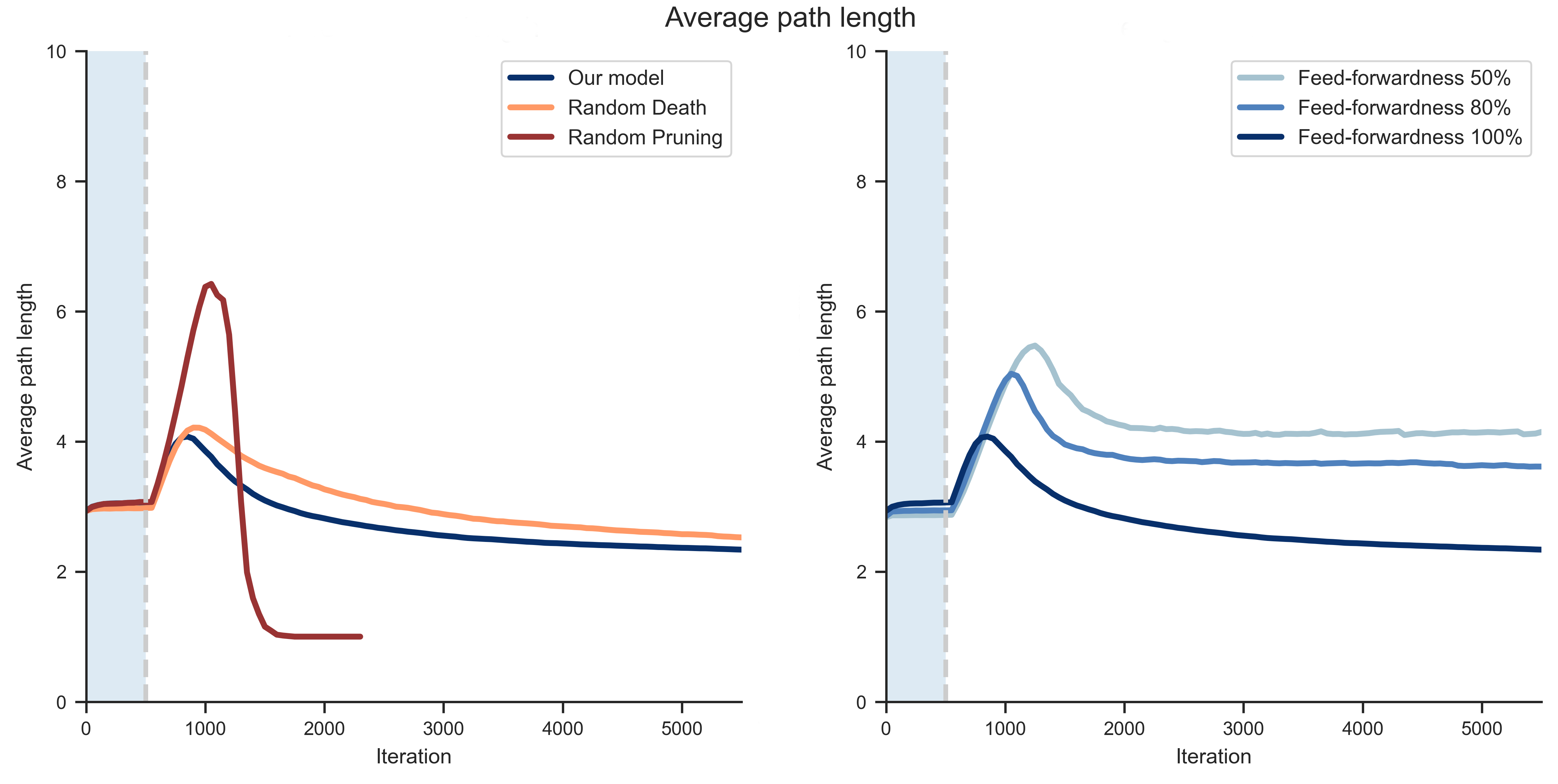}
		\caption{\label{fig:Pathsyn100} Evolution of the average path length in our model. (Left) Comparison with random death (orange) or random pruning (red). (Right) Various shades of blue indicate different degrees of feed-forwardness in the network, i.e. the percentage of regulatory connections. The shaded portion of the graph from 0 to 500 iterations represents the ND stage. }
	\end{center}
	
\end{figure}

When evaluating the complex network statistics of our model, we decided to also test different percentages of regulatory connections in the network. In other words, we varied the level of feed-forwardness of the network additionally. In Figure \ref{fig:Pathsyn100} we can see the evolution of the average shortest path length (ASPL), a measure that can be used as a proxy for the efficiency of information or signal propagation within the network, providing insights into the overall connectivity and navigability of the network structure. During the death stage, all the ASPLs are constant at 2.86. However, during pruning, the ASPL is raised for around 500 iterations, especially in the network that has random pruning, where it gets to 6.41 only to decrease later in all networks but that one (the network with random pruning dies around 1500 iterations). By the end of the simulation, the ALSP returns to a lower level than before the pruning stage at 2.39 (2.52 if the death is random). Here, a critical ingredient seems to be the level of feed-forwardness of the network, apart from having a noticeably erratic behaviour throughout the simulation, for both 50\% and 80\% the ALSP is much higher than for the base model (100\% of feed-forwardness) at 4.52 and 3.61, respectively. This points to a network that is harder to navigate as you add more regulatory connections. 

\begin{figure} [H]
	
	\begin{center}
		\includegraphics[width=0.88\textwidth]{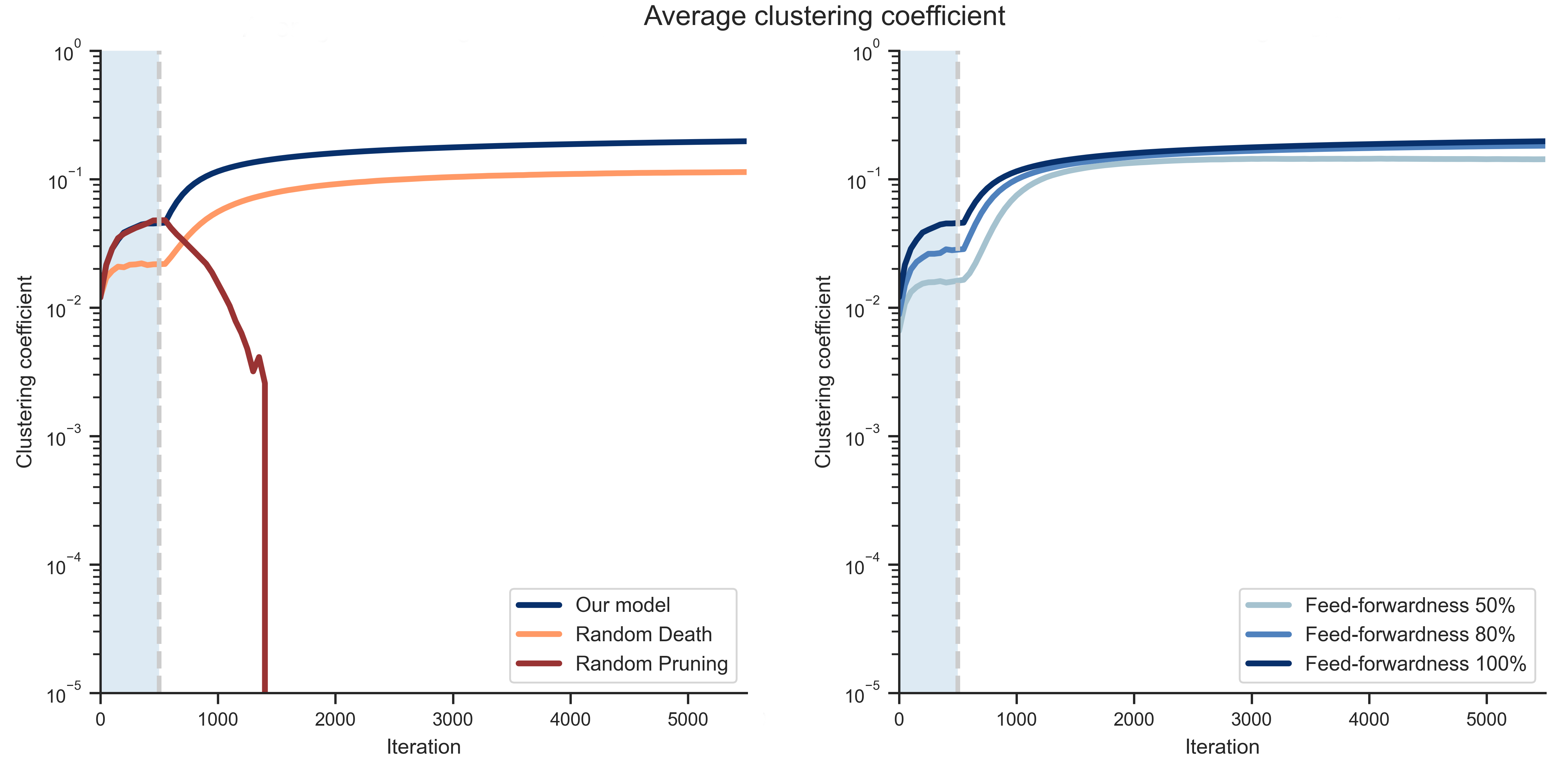}
		\caption{\label{fig:Clusteringsyn100} Evolution of the clustering coefficient in our model. (Left) Comparison with random death (orange) or random pruning (red). (Right) Various shades of blue indicate different degrees of feed-forwardness in the network, i.e. the percentage of regulatory connections. The shaded portion of the graph from 0 to 500 iterations represents the ND stage. }
	\end{center}
	
\end{figure}

Brain networks are also expected to be cohesive and modular. To test this, we evaluated the clustering coefficient of all the network snapshots as time progressed in the simulations (Figure \ref{fig:Clusteringsyn100}).  Here, the key element seems to be the selective pruning once more, without which the network loses all cohesiveness by iteration 1500 (Figure \ref{fig:Clusteringsyn100}, red in the left panel). For all other conditions, there is a sharp increase in the coefficient during the ND stage followed by a stabilisation phase. Notably, there is a marked delay in this stable phase for the lower levels of feed-forwardness (Figure \ref{fig:Clusteringsyn100}, darker shade of blue in the right panel). This figure also shows a critical role for the selective death in the model, by the end of that stage the clustering coefficient is 0.021 if the death is random and 0.045 if it is performed selectively in the way we propose. Both networks seem to have a similar stabilisation stage during the pruning iterations ending up with a clustering coefficient of 0.11 if the death is random and 0.19 if it is selective. The role of the feed-forwardness in this stage is less pronounced, with a clustering coefficient of 0.14 for 50\% and 0.18 for 80\%, remarkbly 
 similar to our base model.

 \subsection{Synaptic preservation and pruning rates}

\begin{figure} [H]
	
	\begin{center}
		\includegraphics[width=0.88\textwidth]{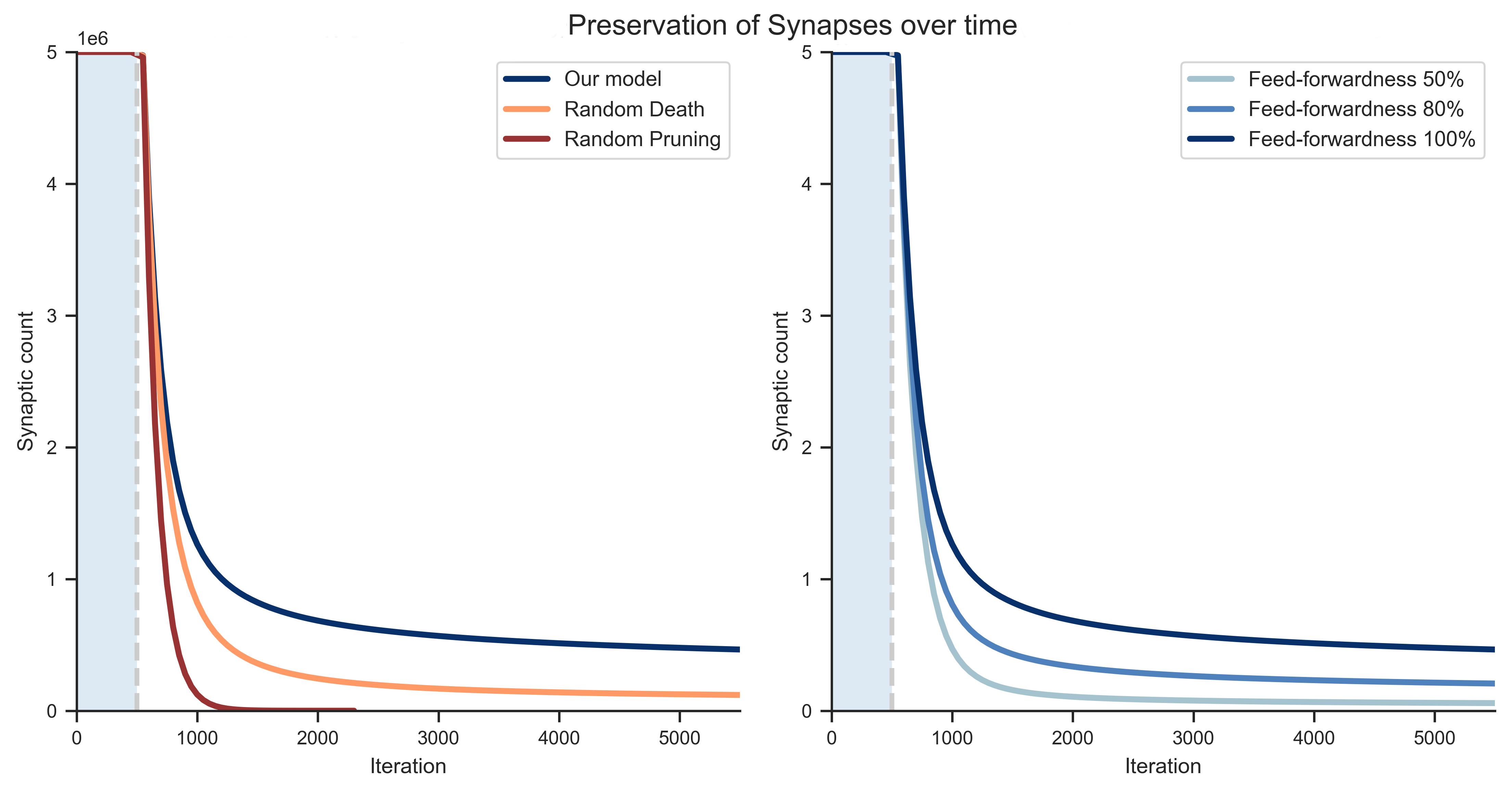}
		\caption{\label{fig:synaptic_preservationsyn100} Number of synapses (edges) in the network as time progresses in the simulation. (Left) Comparison with random death (orange) or random pruning (red). (Right) Various shades of blue indicate different degrees of feed-forwardness in the network, i.e. the percentage of regulatory connections. The shaded portion of the graph from 0 to 500 iterations represents the ND stage.  }
	\end{center}
	
\end{figure}

 The cohesive nature found in figure \ref{fig:Clusteringsyn100} led us to investigate the overall number of connections in the network when our hypothesis is followed (Figure \ref{fig:synaptic_preservationsyn100}). It is worth remembering that this is only possible in the SP stage since the synapses (edges) are re-introduced to the network during the ND stage. Markedly, the biologically-inspired model generates a more densely connected network than the alternatives tested and this seems to be directly proportional to the percentage of feed-forward connections in the network (Figure \ref{fig:synaptic_preservationsyn100}, right panel). By the end of the simulation, our model preserves 9.38\% of the overall number of synapses versus only 2.32\% if the death stage is performed randomly. This explains the increased clustering of our model in comparison to the simulations with random death. Furthermore, our model preserves almost double the number of connections when compared with the feed-forwardness set to 80\%, where we have 4.17\%, and if the feed-forwardness is set as 50\% only 1\% of the synapses remain in the network.

\begin{figure} [H]
	
	\begin{center}
		\includegraphics[width=1\textwidth]{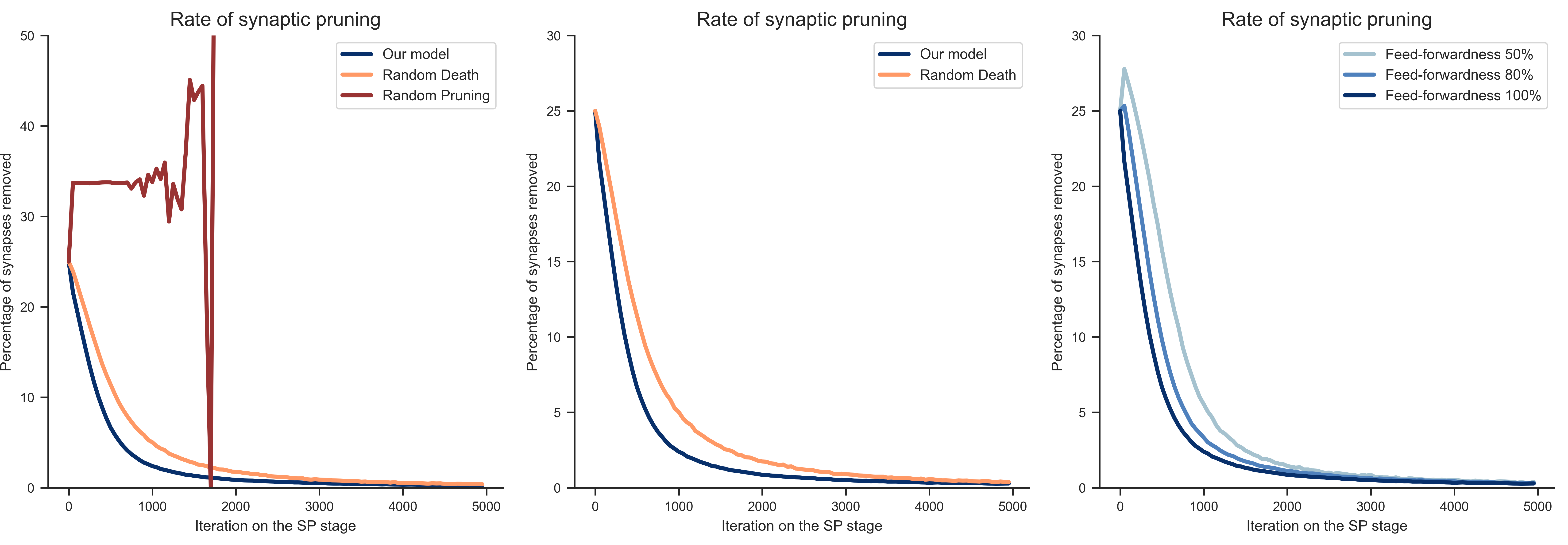}
		\caption{\label{fig:pruning_ratesyn100} Percentage of connections (synapses) removed from the network per iteration of the simulation. (Left) Comparison with random death (orange) or random pruning (red). (Center) A closer look at the rates of synaptic pruning of the random death (orange) vs our model (blue). (Right) Various shades of blue indicate different degrees of feed-forwardness in the network, i.e. the percentage of regulatory connections.}
	\end{center}
	
\end{figure}

Biological neuronal networks were recently found to have a decreasing pruning rate \citep{navlakha2015decreasingRate}. This pattern is also found in our model and it is disrupted when the selective pruning is removed (Figure \ref{fig:pruning_ratesyn100} - red in the left panel). The decrease in the pruning rate is also more pronounced in our model when compared to a network that went through a process of random death (Figure \ref{fig:pruning_ratesyn100} - centre). This sharp decrease in the pruning rate repeats a pattern we found when testing other network statistics -- it seems to be related to the percentage of regulatory connections in the network (Figure \ref{fig:pruning_ratesyn100}, right panel). 

\subsection{Feed-forwardness}

The aforementioned results established our main finding and base model: the biologically inspired dynamics we used to generate a network topology with signs of scale-invariance. Finally, we tested how the percentage of feed-forward synapses affects the evolution of network topology. In the base model, all edges are feed-forward. This is not the case in the cerebral cortex, where feedback and reciprocal connections are common enough such that the complete feed-forward assumption becomes unrealistic \citep{callaway1998local, callaway2004feedforward}. Therefore, we test different percentages of feed-forwardness: 50\% and 80\%. We see that a certain amount of feed-forwardness is a requirement for the topology of the base model - the more, the closer it seems to the original power law, suggesting a causal role for feed-forwardness in the generation of the final network (\textbf{50\%}: D = 0.07, alpha = 2.46; \textbf{80\%}: D = 0.06,  alpha = 2.37; \textbf{100\%}: D = 0.06,  alpha = 2.37, at 5400 iterations).

\begin{figure} [H]
	
	\begin{center}
		\includegraphics[width=1\textwidth]{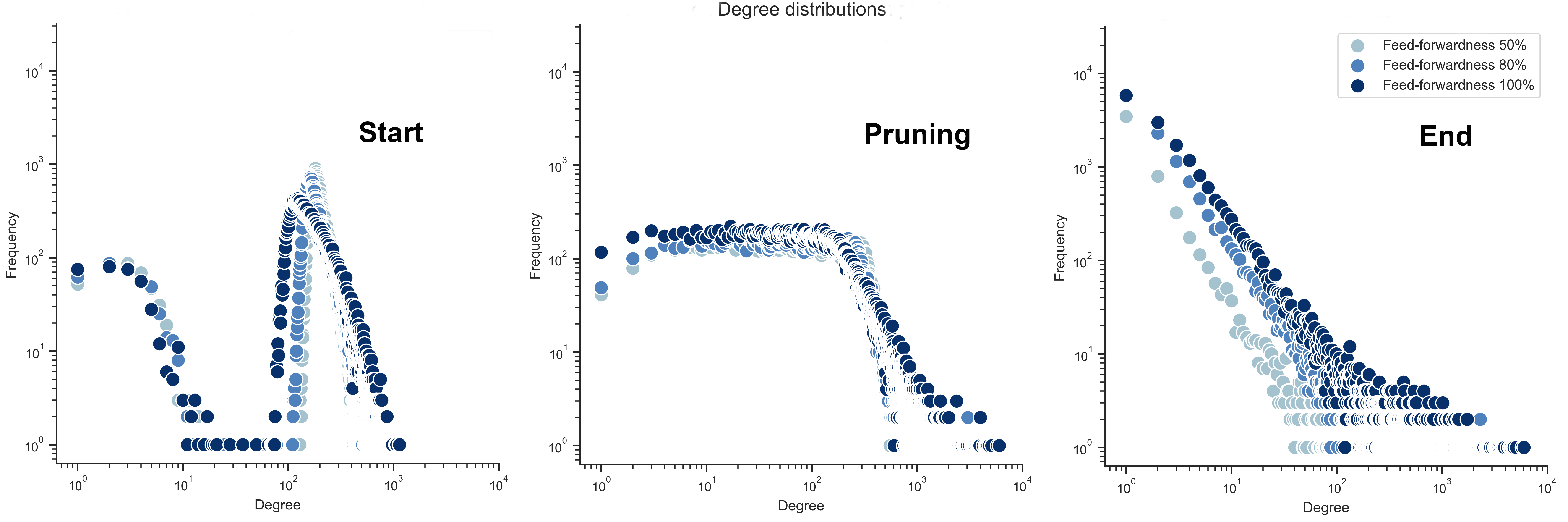}
		\caption{\label{fig:FFsyn100} Distribution of the number of synapses (Degree) of each neuron (Node) in the network. The panels present the progression of the degree distribution in the three stages of our model and shades of blue represent for the different conditions tested, with blue representing different degrees of feed-forwardness in the network.}
	\end{center}
	
\end{figure}

\subsection{Sensitivity analysis of model parameters}

The network development model discussed here, which draws inspiration from brain ontology, primarily revolves around two parameters: A and k, detailed in the Methods section (\ref{Methods}). To assess the resilience and significance of these parameter values, we conducted a robustness test by significantly altering their absolute values. This experiment aimed to determine if minor variations in each parameter would trigger unforeseen model responses or if the model demonstrates adequate robustness to fluctuations in parameter settings. This approach also helps evaluate whether the relationship between the parameter values is crucial for the model to generate the scale-free degree distribution characteristic of heavy-tail connectivity.

\begin{figure} [H]
	
	\begin{center}
		\includegraphics[width=1\textwidth]{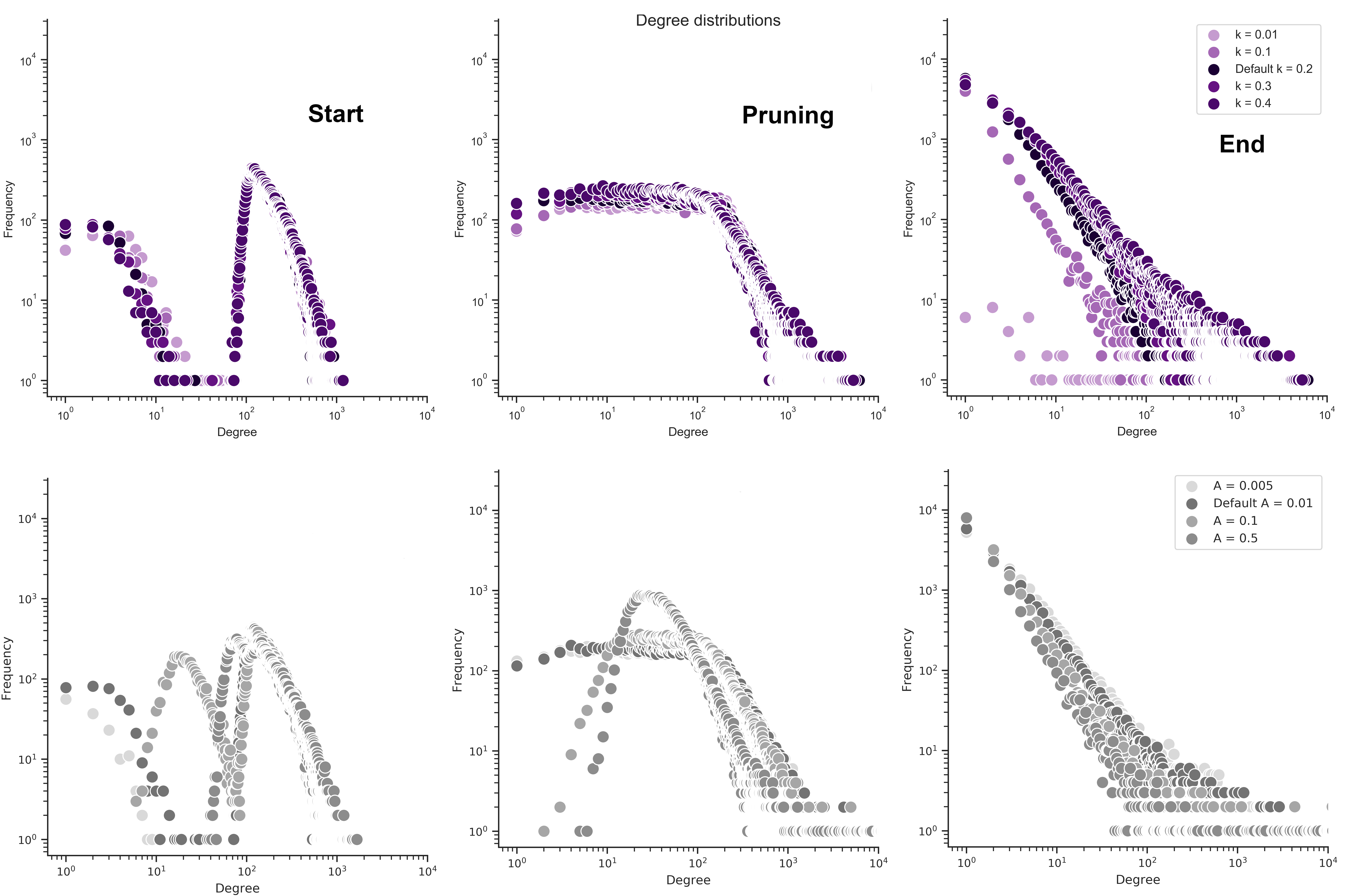}
		\caption{\label{fig:k_and_a} Distribution of the number of synapses (Degree) of each neuron (Node) in the network. The panels present the progression of the degree distribution in our model and shades of colour represent the different conditions tested, with purple representing different numbers of k and grey being used for the variation of A.}
	\end{center}
	
\end{figure}

The aforementioned degree distributions can be seen in figure \ref{fig:k_and_a}. The parameter k appears robust to change (Figure \ref{fig:k_and_a} - Top row) except when lowered to an extreme level (0.01 vs the default 0.2) which, in turn, would render our pruning random and the network would run out of connections as shown in figure \ref{fig:main_result_comparativesyn100}. Lower k values lead to a significant reduction in higher degree nodes, reflecting less selective, more aggressive pruning. Even more robust in terms of power-law generation is the parameter A. All options tested achieved a scale-free degree distribution by the end of the simulation (Figure \ref{fig:k_and_a} - Bottom row), in fact, for the higher end of the variation of A these distributions stabilised as a power law quicker than in the original model.

\begin{figure} [H]
	
	\begin{center}
		\includegraphics[width=0.8\textwidth]{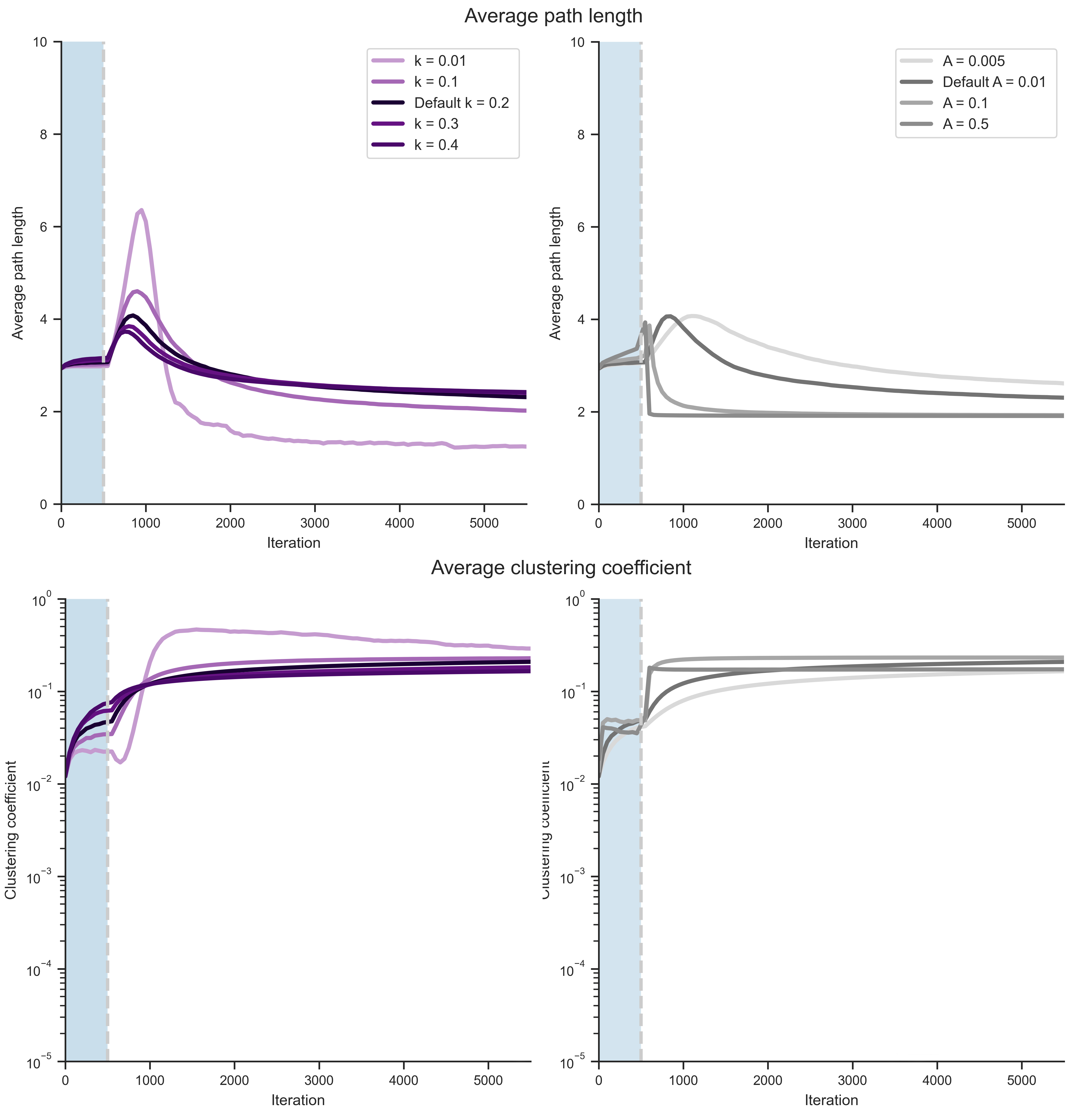}
		\caption{\label{fig:cn_k_and_a} Evolution of the average path length (Top row) and clustering (Bottom row) in our model for different values of the parameters k and A. The panels present the progression of the degree distribution in our model and shades of colour represent the different conditions tested, with purple representing different numbers of k and grey being used for the variation of A.}
	\end{center}
	
\end{figure}

Furthermore, we investigated the effects that the jittering had on the complex network statistics (Figure \ref{fig:cn_k_and_a}). As in the base model case, the ASLP initially increases only to decrease. However, when reducing k to 0.01 we see different behaviour on the ASLP (Figure \ref{fig:cn_k_and_a} - Top left panel) where it becomes quite unstable and does not converge to the same space as the other values, instead of keeping the final ALSP around 2.40  as the other values, it drops it to 1.23 which is consistent with the values found for the networks with random pruning tested against the base model (Figure \ref{fig:Pathsyn100}). The changes in A seem to speed up the decrease of the ALSP to 2 and its stabilisation (Figure \ref{fig:cn_k_and_a} - Top right panel), showing that A is more robust to jittering regarding ALSP than k. Similarly, the parameter A appears more robust to change than k when evaluating the clustering coefficient. For all tested values of k, the clustering coefficient first decreases to its lowest value of 0.01 to then increases to its maximum value of 0.46 to then stabilise at 0.28, slightly higher than the other values that group around 0.2. Finally, high values of the parameter A result in a quick stabilisation of the end values of the clustering coefficients. 

\section{Discussion}

In this study, we utilised the framework provided by network neuroscience to devise a model that suggests that the destructive processes in the ontogeny of the nervous system can lead to heavy-tail connectivity in the form of a scale-invariant neuronal network, which could make it robust to random death of neurons by definition \citep{barabasi2009scale}. This is not to say that these vastly destructive events were exclusively selected for this advantage. One possibility is that it is just a consequence of another necessary event that happens to provide a separate advantage, i.e. a spandrel \citep{gould1979spandrels}. For instance, neurogenesis is heavily concentrated at the beginning of brain development, leading to massive DNA replication errors and aneuploidy \citep{mckinnon2013maintaining}. It could be that the massive apoptosis that follows is a corrective procedure, such that the cells with damaged DNA will not survive long.

Nevertheless, according to our theory, if neuronal death happens as a result of selective detachment, it could serve the purpose of priming the network before the major pruning events. In that way, it will prevent the over-pruning (Figure \ref{fig:pruning_rate}) that can lead not only to the implosion of the network but to known neurological conditions such as schizophrenia \citep{hong2016new, yilmaz2021overexpression, scott2023convergent}. Our model has two main parameters: A, which scales the overall probability of synaptic pruning and neuronal death, and k which regulates the exponential decay based on node connectivity (please refer to equations \ref{eq:death} and \ref{eq:pruning}). The analysis indicates that the model is relatively robust to jittering A, as all tested values resulted in a scale-free degree distribution by the end of the simulation. This suggests that the overall rate of pruning and death can vary without significantly disrupting the network's structural properties. 

In contrast, the parameter k exhibits greater sensitivity. Lower values of k (e.g., k=0.01) lead to a more aggressive and less selective pruning process, resulting in a network that quickly loses its high-degree nodes and deviates from the desired scale-free structure. Higher values of k (e.g., k=0.4), however, maintain a more selective pruning process, preserving the network's integrity and resulting in a more stable power-law distribution. These findings pointed to k having been selected by evolution to ensure effective and selective pruning, according to our theory. It can be that neuronal death was a selected step in brain development that served the purpose of enabling maximum neuronal connectivity with the least amount of individual elements.  

The shift in the power laws seen in the end panel of figure \ref{fig:main_result_comparativesyn100} could also be a consequence of the re-introduction of synapses during the death stage, this could generate some variability in the degree distribution that will go through the pruning stage. In fact, since neuronal death does not seem essential to the model's scale-free behaviour, it would warrant another explanation. One possibility is brought by the results shown in figures \ref{fig:synaptic_preservation} and \ref{fig:pruning_rate}. When neuronal death does not happen selectively, the rate of pruning does not drop optimally, resulting in fewer synapses preserved in the network. Random removal of neurons pushes the initial pruning rate to higher values, leading to a more pronounced reduction in the number of synapses. A similar case of over-pruning is, indeed, seen in schizophrenia \citep{yilmaz2021overexpression}. Conversely, networks generated with random death will still moderately preserve local clustering and connectivity. 

Thus, selective synaptic pruning is our essential ingredient and is a direct consequence of learning \citep{ma2020depletion, scholl2021information}. A recent study measured synaptic loss over time using electronic microscopy images \citep{navlakha2015decreasingRate}. Our model of selective detachment generates the same pattern of decreasing pruning rates (Figure \ref{fig:pruning_rate}). The same study showed that removing edges is an effective way of indirectly designing efficient networks for any purpose \citep{navlakha2015decreasingRate}, lending support to the idea that this is a useful feature for a neuronal network. Our results show that if the reduction of synaptic numbers is performed in a random way, it will result in an unstable and aggressive synaptic removal stage that will quickly decrease the number of connections in the network. Our model seems to produce a controlled and efficient pruning process that maintains network stability preserving local clustering and reducing the ALSP. 

The heavy-tail connectivity presented also would be of behavioural importance as it was recently found in functional connectivity studies \citep{kardan2023improvements}. Kardan \textit{et al.} showed that more efficient information processing during a cognitive task may be indicated by higher fMRI scale-free activity, and this could predict further improvements in task performance \citep{kardan2023improvements}. In terms of complex network statistics, our findings seem to be in line with functional connectivity studies \citep{van2009efficiency}.

Beyond the neurobiological inspirations of the model, it has not escaped our notice that our results have consequences for the theoretical understanding of scale-free networks more generally. Our work suggests that scale-free networks can be generated by a selective deletion mechanism. Previously, all other network models shown to have this characteristic relied on some form of selective attachment \citep{hebert2011structural, bell2017network}. This is, to our knowledge, only the second class of algorithms that generate scale-free networks, the other being the Barabási-Albert/Barabási-Bianconi class of models \citep{barabasiAlbert1999, bianconi2001competition}. While reaching the same scale-free distribution, the algorithms differ in one main aspect: our model is preferential in the deletion of edges while the Barabási-Bianconi model is preferential in the addition of nodes and edges.

The underlying similarities are more interesting, however, as they might suggest a general principle. Taking the simpler Barabási-Albert model as an example, it applies \textit{preferential addition of edges} on top of a \textit{temporal inequality} - older nodes are preferred since they will have spent more iterations able to receive new edges and will likely have accumulated more edges in the course of the simulation. Our model has one required ingredient: selective elimination of edges. However, the resulting network seems more robust when it starts from feed-forward connectivity. This is analogous to the Barabási-Albert model, in the sense that our model applies \textit{preferential deletion of edges} on top of a \textit{spatial inequality}: in a feed-forward network, nodes at the end of the network can receive edges originating in any of the nodes behind it, while nodes at the beginning of the network cannot. As a consequence, end nodes will likely have a higher degree from the beginning.

The possible mechanisms, then, might include not only "richer-gets-richer", as in the Barabási-Bianconi model, but also "poorer-gets-poorer", as in our model. More synthetically, an existing inequality is reinforced by a process: inequality upon inequality. We have scale-free networks more easily when an algorithm that is degree-dependent is applied to a structure that already contains an underlying inequality in degree distribution (temporal or spatial). It remains to be explored how much the particular form of the unequal distribution or the shape of the detachment mechanism matters.

For both deletion and attachment, the various possible definitions of selectiveness define whole classes of models. Presently, we have used a simplified measure of connectivity as a selection function; it is relatively straightforward to incorporate into this function more complex algebraic quantifications of connectivity, as well as wiring and metabolic costs. Alternatively, if one includes any class of simulated spiking activity into the networks, one can preferentially delete neurons/synapses that are the most task- or representation-irrelevant. Indeed, a version of this mechanism has been shown to overcome catastrophic forgetting in artificial networks \citep{peng2021overcoming}. Additionally, scale-free structural connectivity was already demonstrated in the mouse visual cortex's response to natural images when considering the distribution of activity covariances \citep{stringer2019high}. We intend to explore all these alternative implementations of selectiveness in the future. Again, we stress that we are not attempting to simulate cortical development in detail here. Rather, we show that mechanisms and structures known to exist in the developing cortical are sufficient to generate scale-free networks with the characteristics listed above robustly. For now, it was worthwhile to show that scale-invariance can arise in feed-forward networks from the simplest possible version of the selective deletion mechanism.

\section*{Acknowledgements}

The authors would like to thank our dear friend Roberto Marinho who sadly could not see this work all the way to its publication. We would also like to thank Dr Jon Stammers at the AMRC for encouraging Rodrigo's pursuit of his ideas and Dr Hannes Saal at the Active Touch Lab for allowing us to run the simulation and analysis at his group's HPC. We also thank Suzana Herculano-Houzel for the fruitful discussions in the conception of this project. Francesca Redmore and the metaBIO team also deserve thanks for the help with proofreading. 

\section*{Data and code availability}

Data from the model simulations are available at Zenodo. All code for the model and analysis is available on \href{https://github.com/rodrigokazu/bio_inspired_complexnets}{GitHub.}

\bibliographystyle{unsrtnat}
\bibliography{main}  

\begin{thebibliography}{82}
\providecommand{\natexlab}[1]{#1}
\providecommand{\url}[1]{\texttt{#1}}
\expandafter\ifx\csname urlstyle\endcsname\relax
  \providecommand{\doi}[1]{doi: #1}\else
  \providecommand{\doi}{doi: \begingroup \urlstyle{rm}\Url}\fi

\bibitem[Abitz et~al.(2007)Abitz, Nielsen, Jones, Laursen, Graem, and
  Pakkenberg]{abitz2007excess}
Maja Abitz, Rune~Damgaard Nielsen, Edward~G Jones, Henning Laursen, Niels
  Graem, and Bente Pakkenberg.
\newblock Excess of neurons in the human newborn mediodorsal thalamus compared
  with that of the adult.
\newblock \emph{Cerebral Cortex}, 17\penalty0 (11):\penalty0 2573--2578, 2007.

\bibitem[Petanjek et~al.(2011)Petanjek, Juda{\v{s}}, {\v{S}}imi{\'c},
  Ra{\v{s}}in, Uylings, Rakic, and Kostovi{\'c}]{petanjek2011extraordinary}
Zdravko Petanjek, Milo{\v{s}} Juda{\v{s}}, Goran {\v{S}}imi{\'c}, Mladen~Roko
  Ra{\v{s}}in, Harry~BM Uylings, Pasko Rakic, and Ivica Kostovi{\'c}.
\newblock Extraordinary neoteny of synaptic spines in the human prefrontal
  cortex.
\newblock \emph{Proceedings of the National Academy of Sciences}, 108\penalty0
  (32):\penalty0 13281--13286, 2011.

\bibitem[Hamburger et~al.(1990)Hamburger, Hamburger, and
  Oppenheim]{hamburger1990naturally}
Viktor Hamburger, Viktor Hamburger, and Ronald~W Oppenheim.
\newblock Naturally occurring neuronal death in vertebrates.
\newblock \emph{Neuroembryology: The Selected Papers}, pages 126--142, 1990.

\bibitem[Changeux and Danchin(1976)]{changeux1976selectivePruning}
Jean-Pierre Changeux and Antoine Danchin.
\newblock Selective stabilisation of developing synapses as a mechanism for the
  specification of neuronal networks.
\newblock \emph{Nature}, 264\penalty0 (5588):\penalty0 705, 1976.

\bibitem[Bandeira et~al.(2009)Bandeira, Lent, and
  Herculano-Houzel]{bandeira2009changing}
Fabiana Bandeira, Roberto Lent, and Suzana Herculano-Houzel.
\newblock Changing numbers of neuronal and non-neuronal cells underlie
  postnatal brain growth in the rat.
\newblock \emph{Proceedings of the National Academy of Sciences}, 106\penalty0
  (33):\penalty0 14108--14113, 2009.

\bibitem[Clarke(1985)]{clarke1985neuronalDeath}
PGH Clarke.
\newblock Neuronal death in the development of the vertebrate nervous system.
\newblock \emph{Trends in Neurosciences}, 8:\penalty0 345--349, 1985.

\bibitem[Haydar et~al.(1999)Haydar, Kuan, Flavell, and
  Rakic]{haydar1999apoptosisBrainSize}
Tarik~F Haydar, Chia-Yi Kuan, Richard~A Flavell, and Pasko Rakic.
\newblock The role of cell death in regulating the size and shape of the
  mammalian forebrain.
\newblock \emph{Cerebral Cortex}, 9\penalty0 (6):\penalty0 621--626, 1999.

\bibitem[Huttenlocher et~al.(1979)]{huttenlocher1979synapticDensity}
Peter~R Huttenlocher et~al.
\newblock Synaptic density in human frontal cortex-developmental changes and
  effects of aging.
\newblock \emph{Brain Res}, 163\penalty0 (2):\penalty0 195--205, 1979.

\bibitem[Stiles and Jernigan(2010)]{stiles2010basics}
Joan Stiles and Terry~L Jernigan.
\newblock The basics of brain development.
\newblock \emph{Neuropsychology review}, 20\penalty0 (4):\penalty0 327--348,
  2010.

\bibitem[Elston et~al.(2009)Elston, Oga, and
  Fujita]{elston2009spinogenesisPruning}
Guy~N Elston, Tomofumi Oga, and Ichiro Fujita.
\newblock Spinogenesis and pruning scales across functional hierarchies.
\newblock \emph{Journal of Neuroscience}, 29\penalty0 (10):\penalty0
  3271--3275, 2009.

\bibitem[Craik and Bialystok(2006)]{craik2006lifespan}
Fergus~IM Craik and Ellen Bialystok.
\newblock Cognition through the lifespan: mechanisms of change.
\newblock \emph{Trends in cognitive sciences}, 10\penalty0 (3):\penalty0
  131--138, 2006.

\bibitem[Snider et~al.(1992)Snider, Elliott, and Yan]{snider1992axotomy}
William~D Snider, Jeffrey~L Elliott, and Qiao Yan.
\newblock Axotomy-induced neuronal death during development.
\newblock \emph{Developmental Neurobiology}, 23\penalty0 (9):\penalty0
  1231--1246, 1992.

\bibitem[Paolicelli et~al.(2011)Paolicelli, Bolasco, Pagani, Maggi, Scianni,
  Panzanelli, Giustetto, Ferreira, Guiducci, Dumas,
  et~al.]{paolicelli2011synaptic}
Rosa~C Paolicelli, Giulia Bolasco, Francesca Pagani, Laura Maggi, Maria
  Scianni, Patrizia Panzanelli, Maurizio Giustetto, Tiago~Alves Ferreira, Eva
  Guiducci, Laura Dumas, et~al.
\newblock Synaptic pruning by microglia is necessary for normal brain
  development.
\newblock \emph{science}, 333\penalty0 (6048):\penalty0 1456--1458, 2011.

\bibitem[Wake et~al.(2009)Wake, Moorhouse, Jinno, Kohsaka, and
  Nabekura]{wake2009resting}
Hiroaki Wake, Andrew~J Moorhouse, Shozo Jinno, Shinichi Kohsaka, and Junichi
  Nabekura.
\newblock Resting microglia directly monitor the functional state of synapses
  in vivo and determine the fate of ischemic terminals.
\newblock \emph{Journal of Neuroscience}, 29\penalty0 (13):\penalty0
  3974--3980, 2009.

\bibitem[Schafer et~al.(2012)Schafer, Lehrman, Kautzman, Koyama, Mardinly,
  Yamasaki, Ransohoff, Greenberg, Barres, and Stevens]{schafer2012microglia}
Dorothy~P Schafer, Emily~K Lehrman, Amanda~G Kautzman, Ryuta Koyama, Alan~R
  Mardinly, Ryo Yamasaki, Richard~M Ransohoff, Michael~E Greenberg, Ben~A
  Barres, and Beth Stevens.
\newblock Microglia sculpt postnatal neural circuits in an activity and
  complement-dependent manner.
\newblock \emph{Neuron}, 74\penalty0 (4):\penalty0 691--705, 2012.

\bibitem[Thomas et~al.(2016)Thomas, Davis, Karmiloff-Smith, Knowland, and
  Charman]{thomas2016over}
Michael~SC Thomas, Rachael Davis, Annette Karmiloff-Smith, Victoria~CP
  Knowland, and Tony Charman.
\newblock The over-pruning hypothesis of autism.
\newblock \emph{Developmental science}, 19\penalty0 (2):\penalty0 284--305,
  2016.

\bibitem[Teter et~al.(2024)Teter, Draeger, Sattler, McQuade, Holmes, Papakis,
  Leng, Boggess, Shin, Nowakowski, et~al.]{teter2024crispri}
Olivia~M Teter, Nina~M Draeger, Sydney~M Sattler, Amanda McQuade, Brandon~B
  Holmes, Vasileios Papakis, Kun Leng, Steven Boggess, David Shin, Tomasz~J
  Nowakowski, et~al.
\newblock Crispri-based screen of autism spectrum disorder risk genes in
  microglia uncovers roles of adnp in microglia endocytosis and uptake of
  synaptic material.
\newblock \emph{bioRxiv}, pages 2024--06, 2024.

\bibitem[Rosen and Stevens(2010)]{rosen2010role}
Allison~M Rosen and Beth Stevens.
\newblock The role of the classical complement cascade in synapse loss during
  development and glaucoma.
\newblock \emph{Inflammation and Retinal Disease: Complement Biology and
  Pathology}, pages 75--93, 2010.

\bibitem[Howell et~al.(2011)Howell, Macalinao, Sousa, Walden, Soto, Kneeland,
  Barbay, King, Marchant, Hibbs, et~al.]{howell2011molecular}
Gareth~R Howell, Danilo~G Macalinao, Gregory~L Sousa, Michael Walden, Ileana
  Soto, Stephen~C Kneeland, Jessica~M Barbay, Benjamin~L King, Jeffrey~K
  Marchant, Matthew Hibbs, et~al.
\newblock Molecular clustering identifies complement and endothelin induction
  as early events in a mouse model of glaucoma.
\newblock \emph{The Journal of clinical investigation}, 121\penalty0
  (4):\penalty0 1429--1444, 2011.

\bibitem[Wilton et~al.(2023)Wilton, Mastro, Heller, Gergits, Willing, Fahey,
  Frouin, Daggett, Gu, Kim, et~al.]{wilton2023microglia}
Daniel~K Wilton, Kevin Mastro, Molly~D Heller, Frederick~W Gergits, Carly~Rose
  Willing, Jaclyn~B Fahey, Arnaud Frouin, Anthony Daggett, Xiaofeng Gu, Yejin~A
  Kim, et~al.
\newblock Microglia and complement mediate early corticostriatal synapse loss
  and cognitive dysfunction in huntington’s disease.
\newblock \emph{Nature Medicine}, 29\penalty0 (11):\penalty0 2866--2884, 2023.

\bibitem[Yilmaz et~al.(2021)Yilmaz, Yalcin, Presumey, Aw, Ma, Whelan, Stevens,
  McCarroll, and Carroll]{yilmaz2021overexpression}
Melis Yilmaz, Esra Yalcin, Jessy Presumey, Ernest Aw, Minghe Ma, Christopher~W
  Whelan, Beth Stevens, Steven~A McCarroll, and Michael~C Carroll.
\newblock Overexpression of schizophrenia susceptibility factor human
  complement c4a promotes excessive synaptic loss and behavioral changes in
  mice.
\newblock \emph{Nature neuroscience}, 24\penalty0 (2):\penalty0 214--224, 2021.

\bibitem[Hong et~al.(2016{\natexlab{a}})Hong, Beja-Glasser, Nfonoyim, Frouin,
  Li, Ramakrishnan, Merry, Shi, Rosenthal, Barres, et~al.]{hong2016complement}
Soyon Hong, Victoria~F Beja-Glasser, Bianca~M Nfonoyim, Arnaud Frouin, Shaomin
  Li, Saranya Ramakrishnan, Katherine~M Merry, Qiaoqiao Shi, Arnon Rosenthal,
  Ben~A Barres, et~al.
\newblock Complement and microglia mediate early synapse loss in alzheimer
  mouse models.
\newblock \emph{Science}, 352\penalty0 (6286):\penalty0 712--716,
  2016{\natexlab{a}}.

\bibitem[Lui et~al.(2016)Lui, Zhang, Makinson, Cahill, Kelley, Huang, Shang,
  Oldham, Martens, Gao, et~al.]{lui2016progranulin}
Hansen Lui, Jiasheng Zhang, Stefanie~R Makinson, Michelle~K Cahill, Kevin~W
  Kelley, Hsin-Yi Huang, Yulei Shang, Michael~C Oldham, Lauren~Herl Martens,
  Fuying Gao, et~al.
\newblock Progranulin deficiency promotes circuit-specific synaptic pruning by
  microglia via complement activation.
\newblock \emph{Cell}, 165\penalty0 (4):\penalty0 921--935, 2016.

\bibitem[Hong et~al.(2016{\natexlab{b}})Hong, Dissing-Olesen, and
  Stevens]{hong2016new}
Soyon Hong, Lasse Dissing-Olesen, and Beth Stevens.
\newblock New insights on the role of microglia in synaptic pruning in health
  and disease.
\newblock \emph{Current opinion in neurobiology}, 36:\penalty0 128--134,
  2016{\natexlab{b}}.

\bibitem[Scott-Hewitt et~al.(2023)Scott-Hewitt, Huang, and
  Stevens]{scott2023convergent}
Nicole Scott-Hewitt, Youtong Huang, and Beth Stevens.
\newblock Convergent mechanisms of microglia-mediated synaptic dysfunction
  contribute to diverse neuropathological conditions.
\newblock \emph{Annals of the New York Academy of Sciences}, 1525\penalty0
  (1):\penalty0 5--27, 2023.

\bibitem[Hebb et~al.(1949)]{hebb1949organization}
Donald~O Hebb et~al.
\newblock The organization of behavior: A neuropsychological theory, 1949.

\bibitem[Bastrikova et~al.(2008)Bastrikova, Gardner, Reece, Jeromin, and
  Dudek]{bastrikova2008pruningHp}
Natalia Bastrikova, Gregory~A Gardner, Jeff~M Reece, Andreas Jeromin, and
  Serena~M Dudek.
\newblock Synapse elimination accompanies functional plasticity in hippocampal
  neurons.
\newblock \emph{Proceedings of the National Academy of Sciences}, 105\penalty0
  (8):\penalty0 3123--3127, 2008.

\bibitem[Zhou et~al.(2004)Zhou, Homma, and Poo]{zhou2004pruningLTD}
Qiang Zhou, Koichi~J Homma, and Mu-ming Poo.
\newblock Shrinkage of dendritic spines associated with long-term depression of
  hippocampal synapses.
\newblock \emph{Neuron}, 44\penalty0 (5):\penalty0 749--757, 2004.

\bibitem[Butts et~al.(2007)Butts, Kanold, and Shatz]{butts2007burstHebb}
Daniel~A Butts, Patrick~O Kanold, and Carla~J Shatz.
\newblock A burst-based “hebbian” learning rule at retinogeniculate
  synapses links retinal waves to activity-dependent refinement.
\newblock \emph{PLoS biology}, 5\penalty0 (3):\penalty0 e61, 2007.

\bibitem[Zhang et~al.(2012)Zhang, Ackman, Xu, and Crair]{zhang2012hebbVisual}
Jiayi Zhang, James~B Ackman, Hong-Ping Xu, and Michael~C Crair.
\newblock Visual map development depends on the temporal pattern of binocular
  activity in mice.
\newblock \emph{Nature neuroscience}, 15\penalty0 (2):\penalty0 298, 2012.

\bibitem[Ghosh et~al.(1994)Ghosh, Carnahan, and
  Greenberg]{ghosh1994requirement}
Anirvan Ghosh, Josette Carnahan, and Michael~E Greenberg.
\newblock Requirement for bdnf in activity-dependent survival of cortical
  neurons.
\newblock \emph{Science}, 263\penalty0 (5153):\penalty0 1618--1623, 1994.

\bibitem[Herculano-Houzel(2009)]{herculano2009human}
Suzana Herculano-Houzel.
\newblock The human brain in numbers: a linearly scaled-up primate brain.
\newblock \emph{Frontiers in human neuroscience}, page~31, 2009.

\bibitem[Raj and Chen(2011)]{raj2011wiring}
Ashish Raj and Yu-hsien Chen.
\newblock The wiring economy principle: connectivity determines anatomy in the
  human brain.
\newblock \emph{PloS one}, 6\penalty0 (9):\penalty0 e14832, 2011.

\bibitem[Bassett and Sporns(2017)]{bassett2017network}
Danielle~S Bassett and Olaf Sporns.
\newblock Network neuroscience.
\newblock \emph{Nature neuroscience}, 20\penalty0 (3):\penalty0 353--364, 2017.

\bibitem[Bullmore and Sporns(2009)]{bullmore2009complex}
Ed~Bullmore and Olaf Sporns.
\newblock Complex brain networks: graph theoretical analysis of structural and
  functional systems.
\newblock \emph{Nature reviews neuroscience}, 10\penalty0 (3):\penalty0
  186--198, 2009.

\bibitem[Bassett and Bullmore(2017)]{bassett2017small}
Danielle~S Bassett and Edward~T Bullmore.
\newblock Small-world brain networks revisited.
\newblock \emph{The Neuroscientist}, 23\penalty0 (5):\penalty0 499--516, 2017.

\bibitem[Van~den Heuvel et~al.(2016)Van~den Heuvel, Bullmore, and
  Sporns]{van2016comparative}
Martijn~P Van~den Heuvel, Edward~T Bullmore, and Olaf Sporns.
\newblock Comparative connectomics.
\newblock \emph{Trends in cognitive sciences}, 20\penalty0 (5):\penalty0
  345--361, 2016.

\bibitem[Puxeddu et~al.(2020)Puxeddu, Faskowitz, Betzel, Petti, Astolfi, and
  Sporns]{puxeddu2020modular}
Maria~Grazia Puxeddu, Joshua Faskowitz, Richard~F Betzel, Manuela Petti, Laura
  Astolfi, and Olaf Sporns.
\newblock The modular organization of brain cortical connectivity across the
  human lifespan.
\newblock \emph{NeuroImage}, 218:\penalty0 116974, 2020.

\bibitem[Crossley et~al.(2014)Crossley, Mechelli, Scott, Carletti, Fox,
  McGuire, and Bullmore]{crossley2014hubs}
Nicolas~A Crossley, Andrea Mechelli, Jessica Scott, Francesco Carletti, Peter~T
  Fox, Philip McGuire, and Edward~T Bullmore.
\newblock The hubs of the human connectome are generally implicated in the
  anatomy of brain disorders.
\newblock \emph{Brain}, 137\penalty0 (8):\penalty0 2382--2395, 2014.

\bibitem[Van Den~Heuvel and Sporns(2011)]{van2011rich}
Martijn~P Van Den~Heuvel and Olaf Sporns.
\newblock Rich-club organization of the human connectome.
\newblock \emph{Journal of Neuroscience}, 31\penalty0 (44):\penalty0
  15775--15786, 2011.

\bibitem[Stanley et~al.(2013)Stanley, Moussa, Paolini, Lyday, Burdette, and
  Laurienti]{stanley2013defining}
Matthew~L Stanley, Malaak~N Moussa, Brielle~M Paolini, Robert~G Lyday,
  Jonathan~H Burdette, and Paul~J Laurienti.
\newblock Defining nodes in complex brain networks.
\newblock \emph{Frontiers in computational neuroscience}, 7:\penalty0 169,
  2013.

\bibitem[Helmstaedter(2013)]{helmstaedter2013cellular}
Moritz Helmstaedter.
\newblock Cellular-resolution connectomics: challenges of dense neural circuit
  reconstruction.
\newblock \emph{Nature methods}, 10\penalty0 (6):\penalty0 501--507, 2013.

\bibitem[Scheffer et~al.(2020)Scheffer, Xu, Januszewski, Lu, Takemura,
  Hayworth, Huang, Shinomiya, Maitlin-Shepard, Berg,
  et~al.]{scheffer2020connectome}
Louis~K Scheffer, C~Shan Xu, Michal Januszewski, Zhiyuan Lu, Shin-ya Takemura,
  Kenneth~J Hayworth, Gary~B Huang, Kazunori Shinomiya, Jeremy Maitlin-Shepard,
  Stuart Berg, et~al.
\newblock A connectome and analysis of the adult drosophila central brain.
\newblock \emph{elife}, 9:\penalty0 e57443, 2020.

\bibitem[Takemura et~al.(2013)Takemura, Bharioke, Lu, Nern, Vitaladevuni,
  Rivlin, Katz, Olbris, Plaza, Winston, et~al.]{takemura2013visual}
Shin-ya Takemura, Arjun Bharioke, Zhiyuan Lu, Aljoscha Nern, Shiv Vitaladevuni,
  Patricia~K Rivlin, William~T Katz, Donald~J Olbris, Stephen~M Plaza, Philip
  Winston, et~al.
\newblock A visual motion detection circuit suggested by drosophila
  connectomics.
\newblock \emph{Nature}, 500\penalty0 (7461):\penalty0 175--181, 2013.

\bibitem[White et~al.(1986)White, Southgate, Thomson, Brenner,
  et~al.]{white1986structure}
John~G White, Eileen Southgate, J~Nichol Thomson, Sydney Brenner, et~al.
\newblock The structure of the nervous system of the nematode caenorhabditis
  elegans.
\newblock \emph{Philos Trans R Soc Lond B Biol Sci}, 314\penalty0
  (1165):\penalty0 1--340, 1986.

\bibitem[Towlson et~al.(2013)Towlson, V{\'e}rtes, Ahnert, Schafer, and
  Bullmore]{towlson2013rich}
Emma~K Towlson, Petra~E V{\'e}rtes, Sebastian~E Ahnert, William~R Schafer, and
  Edward~T Bullmore.
\newblock The rich club of the c. elegans neuronal connectome.
\newblock \emph{Journal of Neuroscience}, 33\penalty0 (15):\penalty0
  6380--6387, 2013.

\bibitem[Randel et~al.(2014)Randel, Asadulina, Bezares-Calder{\'o}n,
  Veraszt{\'o}, Williams, Conzelmann, Shahidi, and
  J{\'e}kely]{randel2014neuronal}
Nadine Randel, Albina Asadulina, Luis~A Bezares-Calder{\'o}n, Csaba
  Veraszt{\'o}, Elizabeth~A Williams, Markus Conzelmann, R{\'e}za Shahidi, and
  G{\'a}sp{\'a}r J{\'e}kely.
\newblock Neuronal connectome of a sensory-motor circuit for visual navigation.
\newblock \emph{elife}, 3:\penalty0 e02730, 2014.

\bibitem[Lynn et~al.(2024)Lynn, Holmes, and Palmer]{lynn2024heavy}
Christopher~W Lynn, Caroline~M Holmes, and Stephanie~E Palmer.
\newblock Heavy-tailed neuronal connectivity arises from hebbian
  self-organization.
\newblock \emph{Nature Physics}, pages 1--8, 2024.

\bibitem[Stringer et~al.(2019)Stringer, Pachitariu, Steinmetz, Carandini, and
  Harris]{stringer2019high}
Carsen Stringer, Marius Pachitariu, Nicholas Steinmetz, Matteo Carandini, and
  Kenneth~D Harris.
\newblock High-dimensional geometry of population responses in visual cortex.
\newblock \emph{Nature}, 571\penalty0 (7765):\penalty0 361--365, 2019.

\bibitem[Barab{\'a}si(2009)]{barabasi2009scale}
Albert-L{\'a}szl{\'o} Barab{\'a}si.
\newblock Scale-free networks: a decade and beyond.
\newblock \emph{science}, 325\penalty0 (5939):\penalty0 412--413, 2009.

\bibitem[Morter{\'a} and Herculano-Houzel(2012)]{mortera2012age}
Priscilla Morter{\'a} and Suzana Herculano-Houzel.
\newblock Age-related neuronal loss in the rat brain starts at the end of
  adolescence.
\newblock \emph{Frontiers in neuroanatomy}, 6:\penalty0 45, 2012.

\bibitem[Barab{\'a}si and Albert(1999)]{barabasiAlbert1999}
Albert-L{\'a}szl{\'o} Barab{\'a}si and R{\'e}ka Albert.
\newblock Emergence of scaling in random networks.
\newblock \emph{science}, 286\penalty0 (5439):\penalty0 509--512, 1999.

\bibitem[Bianconi and Barab{\'a}si(2001)]{bianconi2001competition}
Ginestra Bianconi and A-L Barab{\'a}si.
\newblock Competition and multiscaling in evolving networks.
\newblock \emph{EPL (Europhysics Letters)}, 54\penalty0 (4):\penalty0 436,
  2001.

\bibitem[Bell et~al.(2017)Bell, Perera, Piraveenan, Bliemer, Latty, and
  Reid]{bell2017network}
Michael Bell, Supun Perera, Mahendrarajah Piraveenan, Michiel Bliemer, Tanya
  Latty, and Chris Reid.
\newblock Network growth models: A behavioural basis for attachment
  proportional to fitness.
\newblock \emph{Scientific reports}, 7\penalty0 (1):\penalty0 1--11, 2017.

\bibitem[Nicosia et~al.(2013)Nicosia, Machida, Wilson, Hancock, Konno, Latora,
  and Severini]{nicosia2013co}
Vincenzo Nicosia, Takuya Machida, Richard Wilson, Edwin Hancock, Norio Konno,
  Vito Latora, and Simone Severini.
\newblock Co-evolution of networks and quantum dynamics: a generalization of
  preferential attachment.
\newblock \emph{Journal of Statistical Mechanics: Theory and Experiment},
  2013\penalty0 (08):\penalty0 P08016, 2013.

\bibitem[Douglas et~al.(1989)Douglas, Martin, and
  Whitteridge]{douglas1989canonical}
Rodney~J Douglas, Kevan~AC Martin, and David Whitteridge.
\newblock A canonical microcircuit for neocortex.
\newblock \emph{Neural computation}, 1\penalty0 (4):\penalty0 480--488, 1989.

\bibitem[Meyer et~al.(2010)Meyer, Wimmer, Oberlaender, De~Kock, Sakmann, and
  Helmstaedter]{meyer2010number}
Hanno~S Meyer, Verena~C Wimmer, Marcel Oberlaender, Christiaan~PJ De~Kock, Bert
  Sakmann, and Moritz Helmstaedter.
\newblock Number and laminar distribution of neurons in a thalamocortical
  projection column of rat vibrissal cortex.
\newblock \emph{Cerebral cortex}, 20\penalty0 (10):\penalty0 2277--2286, 2010.

\bibitem[Herculano-Houzel et~al.(2008)Herculano-Houzel, Collins, Wong, Kaas,
  and Lent]{herculano2008basic}
Suzana Herculano-Houzel, Christine~E Collins, Peiyan Wong, Jon~H Kaas, and
  Roberto Lent.
\newblock The basic nonuniformity of the cerebral cortex.
\newblock \emph{Proceedings of the National Academy of Sciences}, 105\penalty0
  (34):\penalty0 12593--12598, 2008.

\bibitem[Gabi et~al.(2010)Gabi, Collins, Wong, Torres, Kaas, and
  Herculano-Houzel]{gabi2010cellular}
Mariana Gabi, Christine~E Collins, Peiyan Wong, Laila~B Torres, Jon~H Kaas, and
  Suzana Herculano-Houzel.
\newblock Cellular scaling rules for the brains of an extended number of
  primate species.
\newblock \emph{Brain Behavior and Evolution}, 76\penalty0 (1):\penalty0
  32--44, 2010.

\bibitem[Wong and Mar{\'\i}n(2019)]{wong2019developmental}
Fong~Kuan Wong and Oscar Mar{\'\i}n.
\newblock Developmental cell death in the cerebral cortex.
\newblock \emph{Annual review of cell and developmental biology}, 35:\penalty0
  523--542, 2019.

\bibitem[Nagappan-Chettiar et~al.(2023)Nagappan-Chettiar, Yasuda,
  Johnson-Venkatesh, and Umemori]{nagappan2023molecular}
Sivapratha Nagappan-Chettiar, Masahiro Yasuda, Erin~M Johnson-Venkatesh, and
  Hisashi Umemori.
\newblock The molecular signals that regulate activity-dependent synapse
  refinement in the brain.
\newblock \emph{Current Opinion in Neurobiology}, 79:\penalty0 102692, 2023.

\bibitem[Tonra et~al.(1998)Tonra, Curtis, Wong, Cliffer, Park, Timmes, Nguyen,
  Lindsay, Acheson, and DiStefano]{tonra1998axotomy}
James~R Tonra, Rory Curtis, Vivien Wong, Kenneth~D Cliffer, John~S Park, Andrew
  Timmes, Trang Nguyen, Ronald~M Lindsay, Ann Acheson, and Peter~S DiStefano.
\newblock Axotomy upregulates the anterograde transport and expression of
  brain-derived neurotrophic factor by sensory neurons.
\newblock \emph{Journal of Neuroscience}, 18\penalty0 (11):\penalty0
  4374--4383, 1998.

\bibitem[von Bartheld and Butowt(2000)]{von2000expression}
Christopher~S von Bartheld and Rafal Butowt.
\newblock Expression of neurotrophin-3 (nt-3) and anterograde axonal transport
  of endogenous nt-3 by retinal ganglion cells in chick embryos.
\newblock \emph{Journal of Neuroscience}, 20\penalty0 (2):\penalty0 736--748,
  2000.

\bibitem[Van~Rossum et~al.(1995)Van~Rossum, Drake, et~al.]{van1995python}
Guido Van~Rossum, Fred~L Drake, et~al.
\newblock \emph{Python reference manual}, volume 111.
\newblock Centrum voor Wiskunde en Informatica Amsterdam, 1995.

\bibitem[Csardi and Nepusz(2006)]{csardi2006igraph}
Gabor Csardi and Tamas Nepusz.
\newblock The igraph software package for complex network research.
\newblock \emph{InterJournal, Complex Systems}, 1695\penalty0 (5):\penalty0
  1--9, 2006.

\bibitem[Bressert(2012)]{bressert2012scipy}
Eli Bressert.
\newblock Scipy and numpy: an overview for developers.
\newblock 2012.

\bibitem[Hunter(2007)]{hunter2007matplotlib}
J.~D. Hunter.
\newblock Matplotlib: A 2d graphics environment.
\newblock \emph{Computing In Science \& Engineering}, 9\penalty0 (3):\penalty0
  90--95, 2007.
\newblock \doi{10.1109/MCSE.2007.55}.

\bibitem[Waskom et~al.(2017)Waskom, Botvinnik, O'Kane, Hobson, Lukauskas,
  Gemperline, Augspurger, Halchenko, Cole, Warmenhoven, De~Ruiter, Pye, Hoyer,
  Vanderplas, Villalba, Kunter, Quintero, Bachant, Martin, Meyer, Miles, Ram,
  Yarkoni, Williams, Evans, Fitzgerald, {, Brian}, Fonnesbeck, Lee, and
  Qalieh]{Waskom2017}
Michael Waskom, Olga Botvinnik, Drew O'Kane, Paul Hobson, Saulius Lukauskas,
  David~C Gemperline, Tom Augspurger, Yaroslav Halchenko, John~B. Cole, Jordi
  Warmenhoven, Julian De~Ruiter, Cameron Pye, Stephan Hoyer, Jake Vanderplas,
  Santi Villalba, Gero Kunter, Eric Quintero, Pete Bachant, Marcel Martin, Kyle
  Meyer, Alistair Miles, Yoav Ram, Tal Yarkoni, Mike~Lee Williams, Constantine
  Evans, Clark Fitzgerald, {, Brian}, Chris Fonnesbeck, Antony Lee, and Adel
  Qalieh.
\newblock mwaskom/seaborn: v0.8.1 (september 2017), 2017.

\bibitem[Clauset et~al.(2009)Clauset, Shalizi, and Newman]{clauset2009powerlaw}
Aaron Clauset, Cosma~Rohilla Shalizi, and Mark~EJ Newman.
\newblock Power-law distributions in empirical data.
\newblock \emph{SIAM review}, 51\penalty0 (4):\penalty0 661--703, 2009.

\bibitem[Cong et~al.(2020)Cong, Chu, Yang, and Pei]{cong2020comprehensible}
Zicun Cong, Lingyang Chu, Yu~Yang, and Jian Pei.
\newblock Comprehensible counterfactual explanation on kolmogorov-smirnov test.
\newblock \emph{arXiv preprint arXiv:2011.01223}, 2020.

\bibitem[Navlakha et~al.(2015)Navlakha, Barth, and
  Bar-Joseph]{navlakha2015decreasingRate}
Saket Navlakha, Alison~L Barth, and Ziv Bar-Joseph.
\newblock Decreasing-rate pruning optimizes the construction of efficient and
  robust distributed networks.
\newblock \emph{PLoS computational biology}, 11\penalty0 (7):\penalty0
  e1004347, 2015.

\bibitem[Callaway(1998)]{callaway1998local}
Edward~M Callaway.
\newblock Local circuits in primary visual cortex of the macaque monkey.
\newblock \emph{Annual review of neuroscience}, 21\penalty0 (1):\penalty0
  47--74, 1998.

\bibitem[Callaway(2004)]{callaway2004feedforward}
Edward~M Callaway.
\newblock Feedforward, feedback and inhibitory connections in primate visual
  cortex.
\newblock \emph{Neural Networks}, 17\penalty0 (5-6):\penalty0 625--632, 2004.

\bibitem[Gould and Lewontin(1979)]{gould1979spandrels}
Stephen~Jay Gould and Richard~C Lewontin.
\newblock The spandrels of san marco and the panglossian paradigm: a critique
  of the adaptationist programme.
\newblock \emph{Proc. R. Soc. Lond. B}, 205\penalty0 (1161):\penalty0 581--598,
  1979.

\bibitem[McKinnon(2013)]{mckinnon2013maintaining}
Peter~J McKinnon.
\newblock Maintaining genome stability in the nervous system.
\newblock \emph{Nature neuroscience}, 16\penalty0 (11):\penalty0 1523--1529,
  2013.

\bibitem[Ma et~al.(2020)Ma, Chen, Cui, Huang, Nehme, Zhang, Li, Wei, Liong,
  Liu, et~al.]{ma2020depletion}
Xiaokuang Ma, Ke~Chen, Yuehua Cui, Guanqun Huang, Antoine Nehme, Le~Zhang,
  Handong Li, Jing Wei, Katerina Liong, Qiang Liu, et~al.
\newblock Depletion of microglia in developing cortical circuits reveals its
  critical role in glutamatergic synapse development, functional connectivity,
  and critical period plasticity.
\newblock \emph{Journal of Neuroscience Research}, 98\penalty0 (10):\penalty0
  1968--1986, 2020.

\bibitem[Scholl et~al.(2021)Scholl, Rule, and Hennig]{scholl2021information}
Carolin Scholl, Michael~E Rule, and Matthias~H Hennig.
\newblock The information theory of developmental pruning: Optimizing global
  network architectures using local synaptic rules.
\newblock \emph{PLoS Computational Biology}, 17\penalty0 (10):\penalty0
  e1009458, 2021.

\bibitem[Kardan et~al.(2023)Kardan, Stier, Layden, Choe, Lyu, Zhang, Beilock,
  Rosenberg, and Berman]{kardan2023improvements}
Omid Kardan, Andrew~J Stier, Elliot~A Layden, Kyoung~Whan Choe, Muxuan Lyu,
  Xihan Zhang, Sian~L Beilock, Monica~D Rosenberg, and Marc~G Berman.
\newblock Improvements in task performance after practice are associated with
  scale-free dynamics of brain activity.
\newblock \emph{Network Neuroscience}, pages 1--63, 2023.

\bibitem[Van Den~Heuvel et~al.(2009)Van Den~Heuvel, Stam, Kahn, and
  Pol]{van2009efficiency}
Martijn~P Van Den~Heuvel, Cornelis~J Stam, Ren{\'e}~S Kahn, and Hilleke
  E~Hulshoff Pol.
\newblock Efficiency of functional brain networks and intellectual performance.
\newblock \emph{Journal of Neuroscience}, 29\penalty0 (23):\penalty0
  7619--7624, 2009.

\bibitem[H{\'e}bert-Dufresne et~al.(2011)H{\'e}bert-Dufresne, Allard, Marceau,
  No{\"e}l, and Dub{\'e}]{hebert2011structural}
Laurent H{\'e}bert-Dufresne, Antoine Allard, Vincent Marceau, Pierre-Andr{\'e}
  No{\"e}l, and Louis~J Dub{\'e}.
\newblock Structural preferential attachment: Network organization beyond the
  link.
\newblock \emph{Physical review letters}, 107\penalty0 (15):\penalty0 158702,
  2011.

\bibitem[Peng et~al.(2021)Peng, Tang, Jiang, Li, Lei, Lin, and
  Li]{peng2021overcoming}
Jian Peng, Bo~Tang, Hao Jiang, Zhuo Li, Yinjie Lei, Tao Lin, and Haifeng Li.
\newblock Overcoming long-term catastrophic forgetting through adversarial
  neural pruning and synaptic consolidation.
\newblock \emph{IEEE Transactions on Neural Networks and Learning Systems},
  33\penalty0 (9):\penalty0 4243--4256, 2021.

\end{thebibliography}

\section*{Appendix: Sparse Networks}

This appendix explores the structure of sparse neural networks with respect to scale-free properties. That is, we studied networks having 10 synapses on average per neuron, which results in a total of 500,000 connections, significantly less dense than the five million synapses that were analysed in the main findings section. Besides, these simulations have an overall 3000 iterations in opposition to the main text, which runs for 5500 for the network as the network is pruned 5000 times.

\begin{figure} [H]
	
	\begin{center}
		\includegraphics[width=1\textwidth]{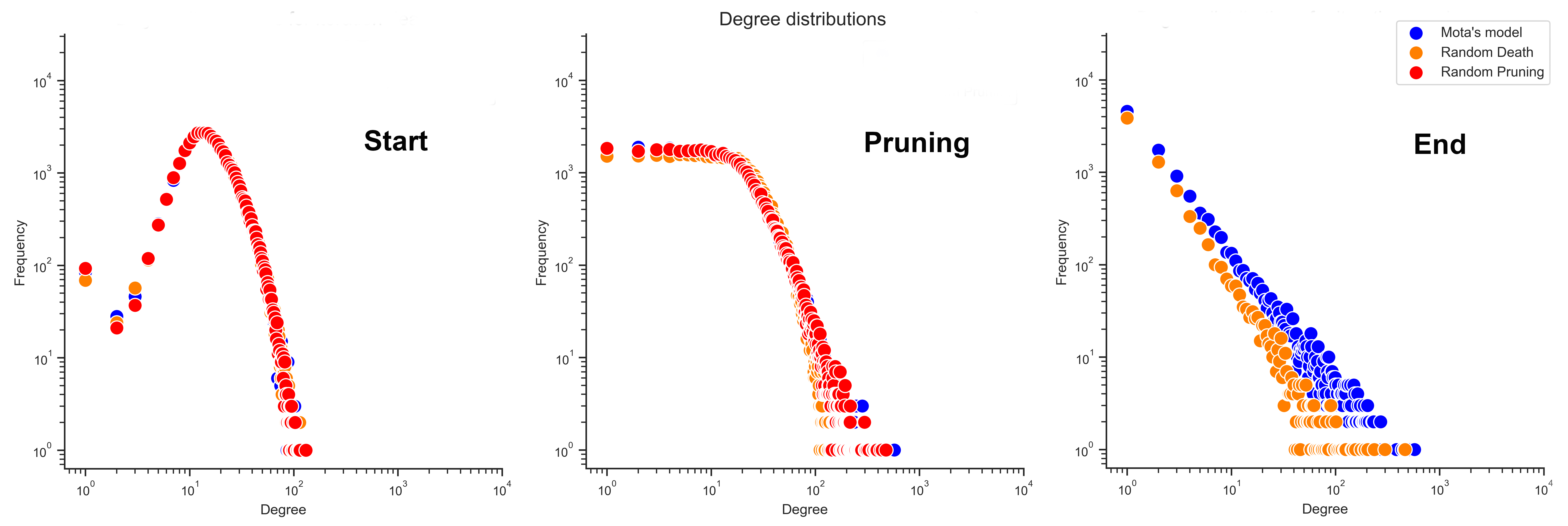}
		\caption{\label{fig:main_result_comparative} Distribution of the number of synapses (degree) of each neuron (node) in the network. The panels present the progression of the degree distribution in our model and colours stand for the different conditions tested, with blue being our base model with selective death and pruning, orange being when only the death is random and red when only the pruning is random.}
	\end{center}
	
\end{figure}

\begin{figure} [H]
	
 	\begin{center}
 		\includegraphics[width=1\textwidth]{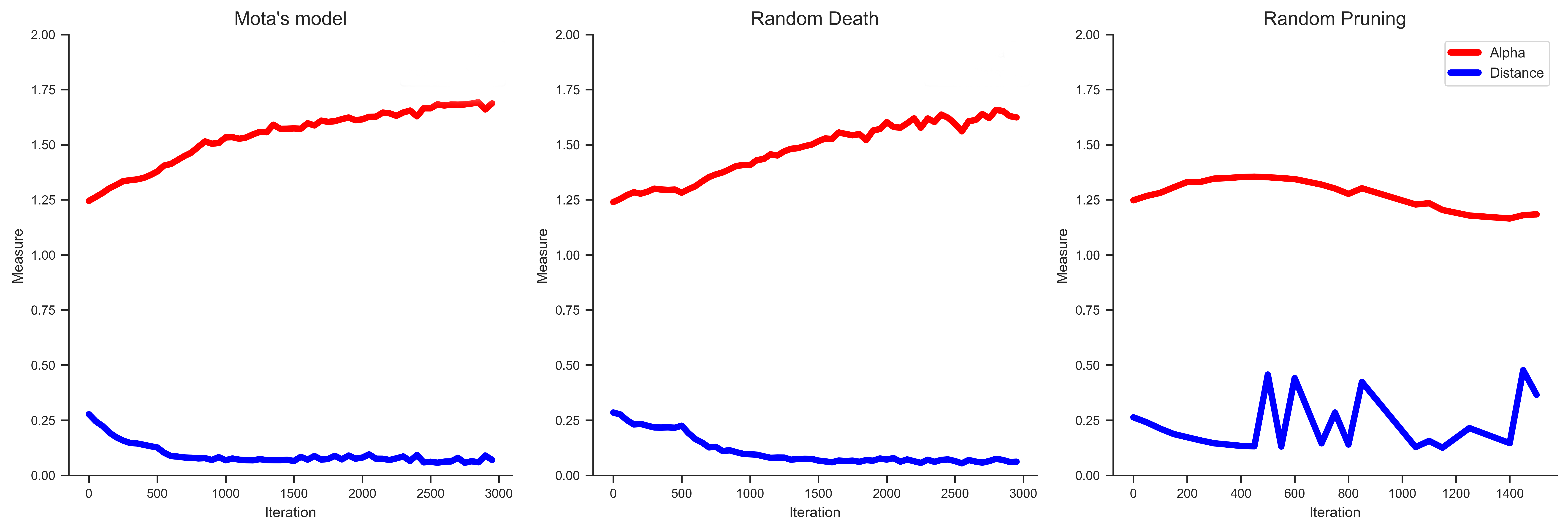}
 		\caption{\label{fig:Alpha_D} The evolution of the exponent alpha and the Kolmogorov distance D as the simulation evolves.}
 	\end{center}
	
 \end{figure}

Figure \ref{fig:main_result_comparative} shows degree distributions revealing heavy-tailed features after selective death and pruning (blue) (K-S test can be seen in figure \ref{fig:Alpha_D}). This is different from what happens during random neuronal death (orange) or after random synaptic pruning (red), which are also not well organized but don’t exhibit power-law characteristics. Therefore, even when densities reduce for sparser structures, specific mechanisms used to prune must be retained to maintain scale-free topology.

\begin{figure} [H]
	
	\begin{center}
		\includegraphics[width=0.77\textwidth]{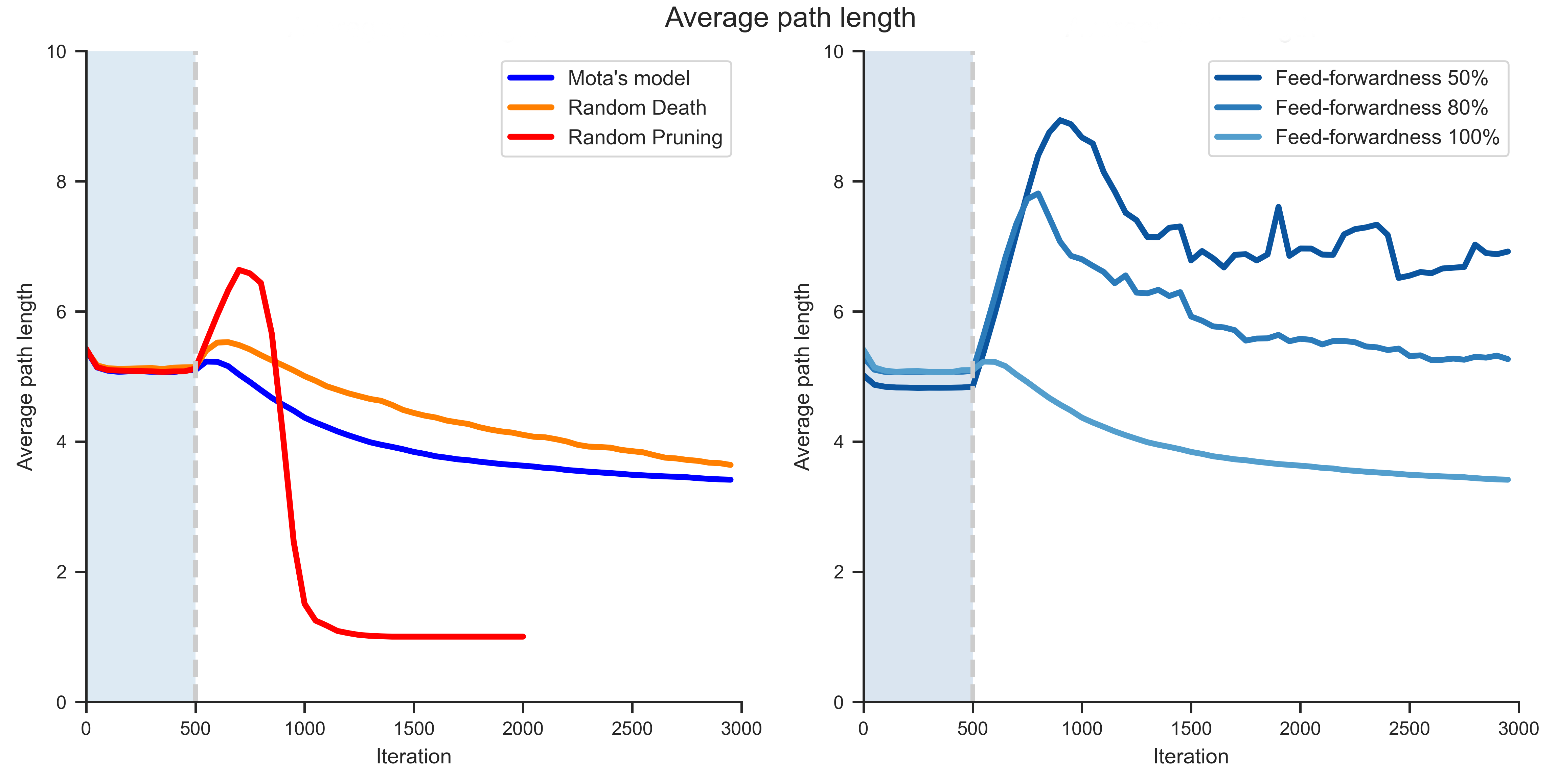}
		\caption{\label{fig:Path} Evolution of the average path length in our model. (Left) Comparison with random death (orange) or random pruning (red). (Right) Various shades of blue indicate different degrees of feed-forwardness in the network, i.e. the percentage of regulatory connections. The shaded portion of the graph from 0 to 500 iterations represents the ND stage. }
	\end{center}
	
\end{figure}

\begin{figure} [H]
	
	\begin{center}
		\includegraphics[width=0.77\textwidth]{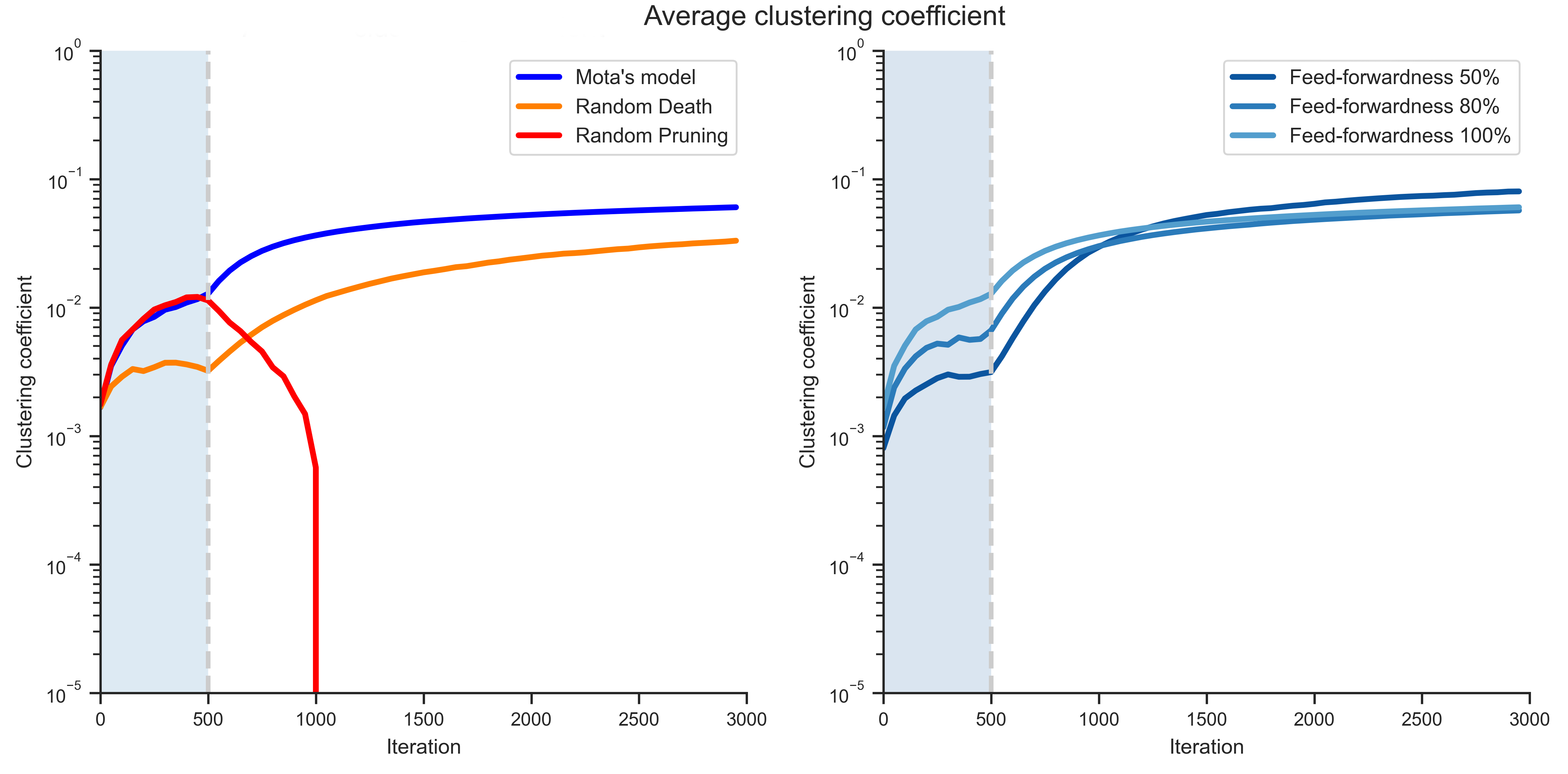}
		\caption{\label{fig:Clustering} Evolution of the clustering coefficient in our model. (Left) Comparison with random death (orange) or random pruning (red). (Right) Various shades of blue indicate different degrees of feed-forwardness in the network, i.e. the percentage of regulatory connections. The shaded portion of the graph from 0 to 500 iterations represents the ND stage. }
	\end{center}
	
\end{figure}

\begin{figure} [H]
	
	\begin{center}
		\includegraphics[width=0.77\textwidth]{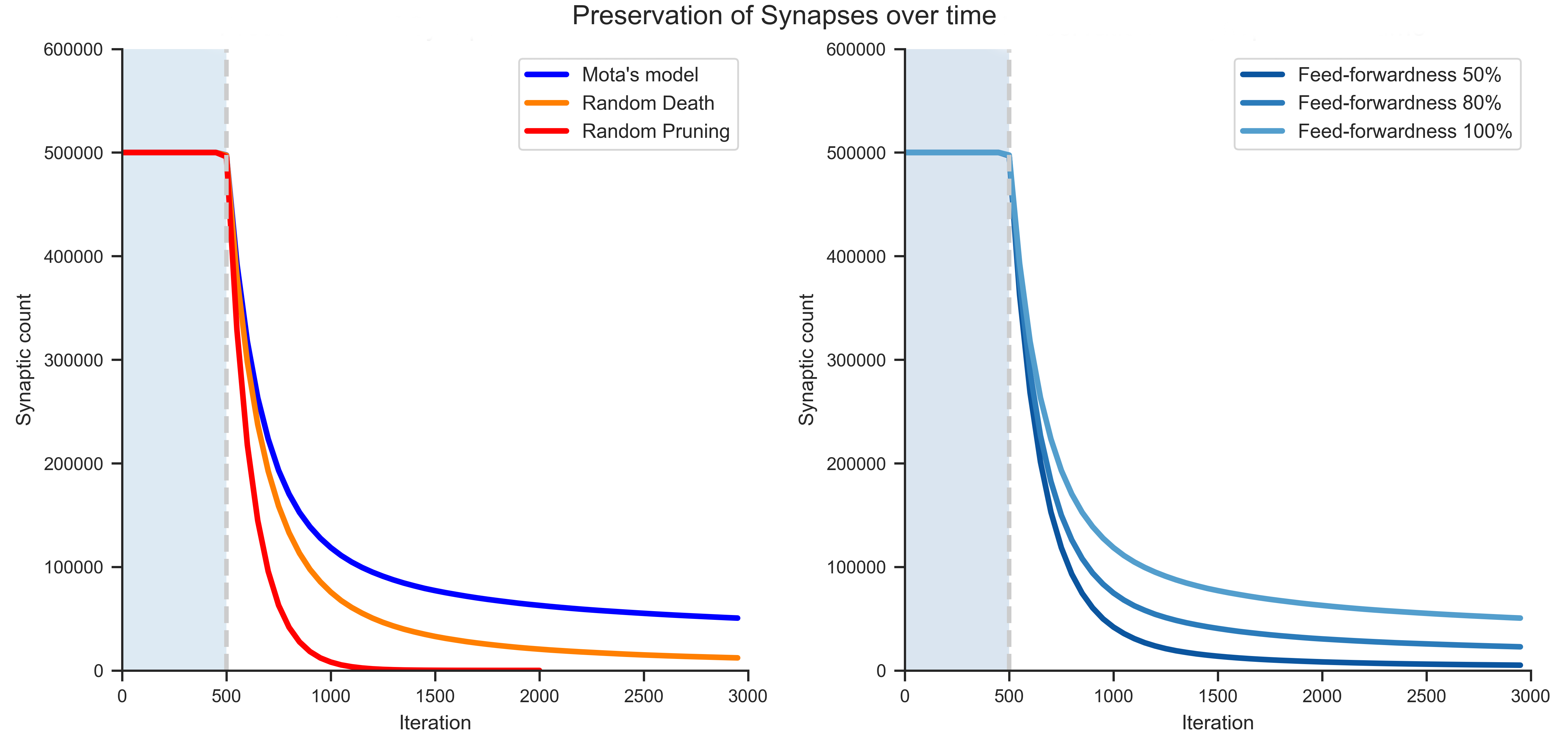}
		\caption{\label{fig:synaptic_preservation} Number of synapses (edges) in the network as time progresses in the simulation. (Left) Comparison with random death (orange) or random pruning (red). (Right) Various shades of blue indicate different degrees of feed-forwardness in the network, i.e. the percentage of regulatory connections. The shaded portion of the graph from 0 to 500 iterations represents the ND stage.  }
	\end{center}
	
\end{figure}

\begin{figure} [H]
	
	\begin{center}
		\includegraphics[width=1\textwidth]{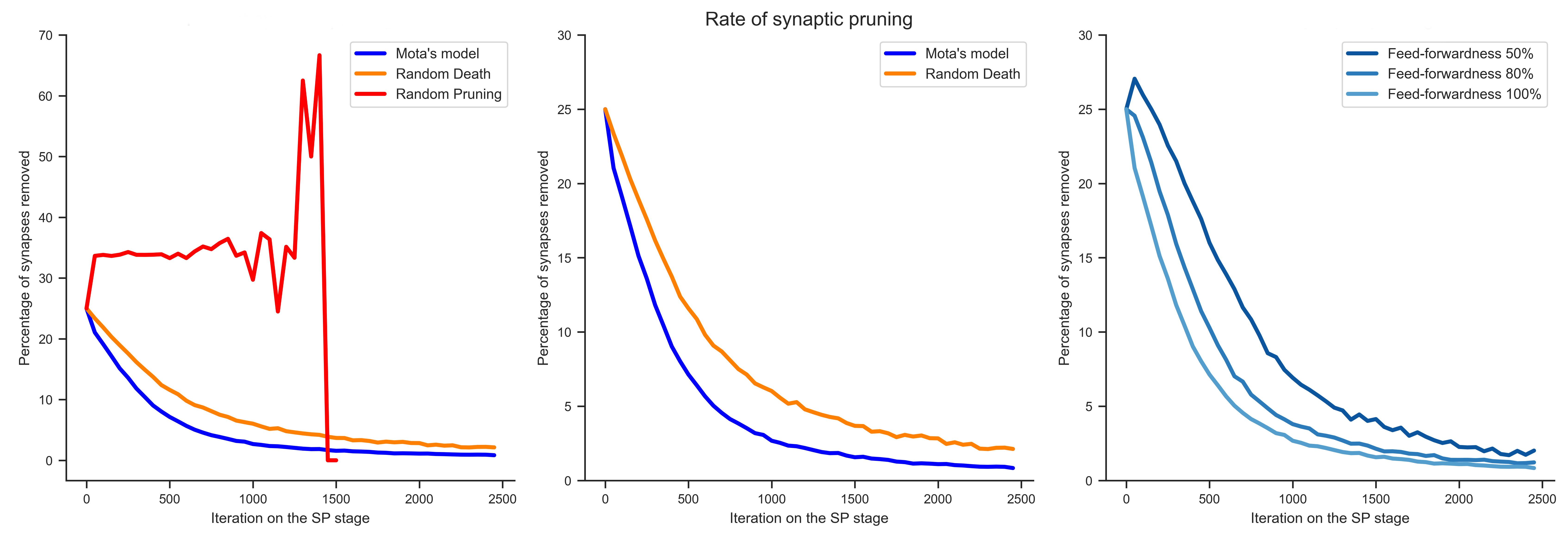}
		\caption{\label{fig:pruning_rate} Percentage of connections (synapses) removed from the network per iteration of the simulation. (Left) Comparison with random death (orange) or random pruning (red). (Center) A closer look at the rates of synaptic pruning of the random death (orange) vs our model (blue). (Right) Various shades of blue indicate different degrees of feed-forwardness in the network, i.e. the percentage of regulatory connections.}
	\end{center}
	
\end{figure}

Figures 12 – 17 illustrate the evolution of network metrics over time. The K-S distance and power-law exponent (Figure \ref{fig:Alpha_D}) indicate that the network increasingly obeys power-law distribution. At 2350 iterations, for instance, the K-S distance (D) is measured at 0.059 with an alpha exponent equal to 1.693 which is indicative of a stable power-law distribution. Average path length (Figure \ref{fig:Path}) and clustering coefficient (Figure \ref{fig:Clustering}) show that selective pruning maintains network efficiency and cohesiveness, whereas random pruning disrupts these properties. Figure \ref{fig:synaptic_preservation} illustrates the number of synapses preserved whilst the pruning rates are shown in figure \ref{fig:pruning_rate}. A high number of feed-forward connections in networks ensures more controlled and steady pruning when compared to low feed-forward connection networks that lead to the maintenance of functionality as well as structure.

Contrary to the main results with denser networks and longer iterations, in sparse networks selective pruning and neuron death have a higher impact on preservation of network properties. The overall iteration time is shorter—3000 iterations—hence the network takes less time to stabilize leading to higher differences in network metrics between selective and random pruning conditions. Sparse networks are more prone to loss of their structural integrity than dense ones. 

\begin{figure} [H]
	
	\begin{center}
		\includegraphics[width=1\textwidth]{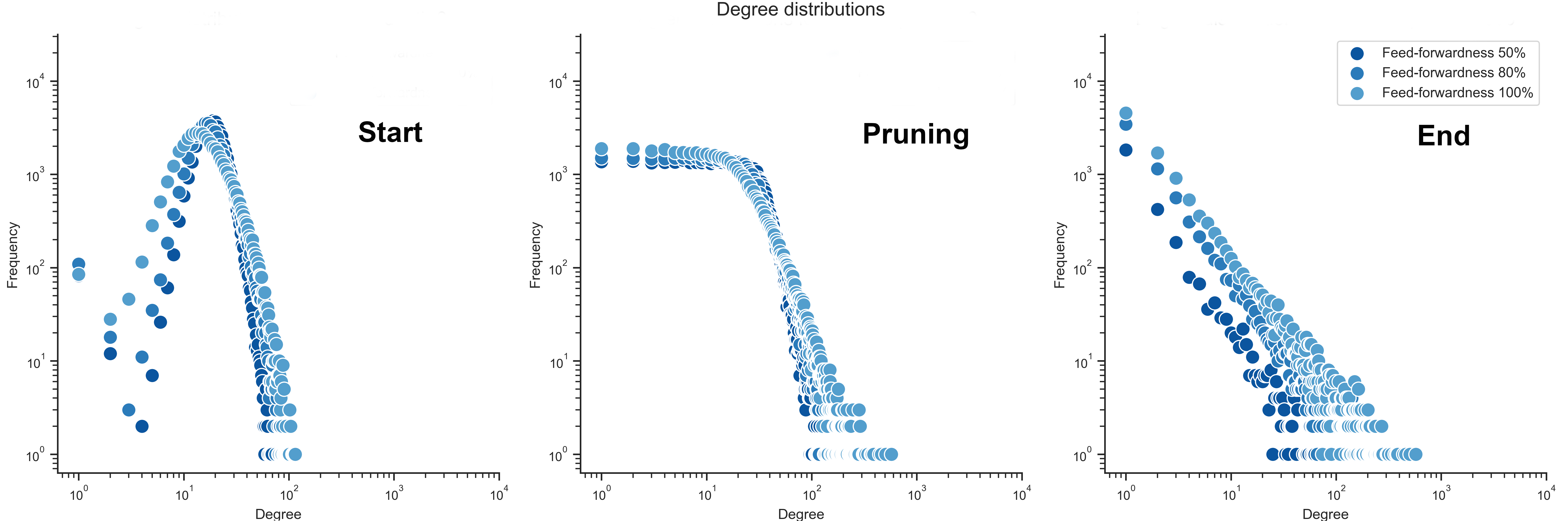}
		\caption{\label{fig:FF} Distribution of the number of synapses (Degree) of each neuron (Node) in the network. The panels present the progression of the degree distribution in our model and shades of blue represent the different conditions tested, with blue representing different degrees of feed-forwardness in the network, i.e. the percentage of regulatory connections of 50\% and 80\%.}
	\end{center}
	
\end{figure}

Finally, Figure \ref{fig:FF} shows that the degree of feed-forwardness affects the evolution of a network greatly. For instance, a high number of feed-forward connections leads to distribution that is much like that in the case of the base model, thus confirming an assertion made earlier about the significance of these feed-forward links in building scale-free networks. On the other hand, decreased levels of such connections have implications for the power law behaviour, suggesting a causal role for feed-forwardness in the generation of the final network (\textbf{50\%}: D = 0.073, alpha = 1.827, at 2350 iterations; \textbf{80\%}: D = 0.077,  alpha = 1.692, at 2350 iterations; \textbf{100\%}: D = 0.059,  alpha = 1.693, at 2350 iterations).

Thus selective pruning and neuronal death play a vital role in constructing robust highly efficient scale-free neuron networks especially sparse ones with short evolution times; this relates to the findings above concerning development time as well as measures taken during the postnatal period. These observations not only provide insight into brain development but also disclose valuable ideas about network science suggesting that specific disconnecting mechanisms apply widely to building scale-free networks across different fields particularly when density is low and the time required for evolution is short-lived.

\end{document}